\documentclass[pra,preprint,superscriptaddress,showpacs,preprintnumbers,amsmath,amssymb,eqsecnum,]{revtex4}

\usepackage{amsmath,amssymb}
\usepackage[dvipsnames]{xcolor}

\usepackage{epsfig}    
\usepackage{dcolumn}
\usepackage{bm}
\usepackage{soul}
\usepackage{makeidx}

\makeindex

\definecolor{darkgreen}{rgb}{0,0.6,0}

\begin{document}

\title{
The Uncertainty Principle Revisited
}

\author{Ady Mann}
\address{Physics Department, Technion - Israel Institute of Technology, 
Haifa 32000, Israel}

\author{Pier A. Mello}
\address{Instituto de F\'isica, Universidad Nacional Aut\'onoma de 
M\'exico, 04510, Cd. de M\'exico, Mexico}

\author{Michael Revzen}
\address{Physics Department, Technion - Israel Institute of Technology, 
Haifa 32000, Israel}
\begin{abstract} 
We study the quantum-mechanical uncertainty relation originating from the successive measurement of two observables
$\hat{A}$ and $\hat{B}$, 
with eigenvalues $a_n$ and $b_m$, respectively,                                            
performed on the same system.
We use an extension of the von Neumann model of measurement, in which two probes interact with the same system proper at two successive times, so we can exhibit how the disturbing effect of the first interaction affects the second measurement.
Detecting the statistical properties  of the second {\em probe} variable $Q_2$ conditioned on the first {\em probe} measurement yielding $Q_1$ we obtain information on the statistical distribution
of the {\em system} variable $b_m$ conditioned on having found the system variable $a_n$ 
in the interval 
$\delta a$ around $a^{(n)}$.
The width of this statistical distribution as function of $\delta a$ constitutes an 
{\em uncertainty relation}.
We find a general connection of this uncertainty relation with the commutator of the two 
observables that have been measured successively.
We illustrate this relation for the successive measurement of position and momentum in the discrete and in the continuous cases and, within a model, for the successive measurement of a more general class of observables.
\end{abstract}



\maketitle

\section{Introduction}
\label{intro}

\index{Introduction}

The uncertainty principle (UP), broadly speaking, expresses interrelation among non-commeasurable physical attributes.
Phrased mathematically, the principle is given in terms of uncertainty relations (UR) establishing
correlations among such non-commuting observables
\cite{vN}.


A convenient explicit illustration for the perhaps somewhat vague statement above, due in the main to the rather unusual correlations involved, will now be given for
the prime example of such quantities, position and momentum. Here the principle is accounted for essentially in two forms. 

The first, to be termed ``{\em preparation}" form, constrains the values of the standard deviations (i.e. the uncertainties) of the
non-commeasurable attributes $\hat{x}$ (position) and $\hat{p}$ (momentum). 
Within the formal framework of Quantum Mechanics, the uncertainty relation acquires the familiar form:
given a system described by a Quantum Mechanical state at time $t$, the standard deviation of the position over an ensemble of systems at time $t$, times the standard deviation of momentum over an independent ensemble of equally prepared systems, also at time $t$, cannot be smaller than $\hbar/2$.
Note that each sample is subjected to one measurement only.
About this formal statement, we quote A. Peres in his book, 
Ref. [\onlinecite{peres_book}], p. 93:
``This is not a statement about the accuracy of our measuring instruments...
There never is any question here that a measurement of $x$ ``disturbs" the value of $p$ 
and vice-versa, as sometimes claimed.
These measurements ... are performed on {\em different} particles ... and therefore these measurement cannot disturb each other in any way."
The uncertainties appear as intrinsic properties, 
not related to the disturbance produced by measurements.


The second form of the UP, to be termed ``{\em measurement}" form, involves the disturbance induced by the measurement on the state:
therefore, in a sequence of non-commeasurable measurements, the second measurement relates to a disturbed state. 
Thus a relation between these two forms, though intuitively suggestive, is not 
at all obvious. 

The second form is perhaps closer to Heisenberg's  original formulation of the UR,
$\Delta x \Delta p \sim h$, using a $\gamma$-ray microscope:
Heisenberg \cite{heisenberg_1927} suggested that the 
disturbance produced by the measurement was the source of the uncertainty.
Indeed, according to Ref. [\onlinecite{sudbery}], p. 25,
``Heisenberg originally explained the UP in terms of the uncontrollable 
change in momentum which is caused by determining the particle's position, 
$\cdots$".
However, it is interesting to notice that, in the same article,
Ref. \cite{heisenberg_1927}, Heisenberg derived the uncertainty relation from an elementary analysis of wave properties, based on Schr\"odinger's wave-mechanical views,
this being the first form described above.
We may also mention that in his lecture notes on 
{\em The Physical Principles of the Quantum Theory} \cite{heisenberg_1949},  Heisenberg presented his UR
as a property of wave packets on pp. 13-15, with a treatment similar to that of Robertson's [\onlinecite{robertson}] on pp. 15-18,
and with his famous $\gamma$-ray microscope on p. 21.


There are numerous studies (\cite{vN,B1,arthurs_kelly,B2,Ent} and references therein) based on von Neumann's
measurement model that relate the two forms. 
Nonetheless, it seems to us that no definitive general relation
between the two is available. 

In a rather recent paper, Ref. \cite{B1}, 
the authors discuss the various aspects of the UR:
i) The role of preparation, which corresponds to the formal statement within the QM formalism, as explained above.
ii) The role of simultaneous measurements.
This aspect has been treated by 
Arthurs and Kelly \cite{arthurs_kelly} and is summarized by the authors of Ref. \cite{B1}.
Arthurs and Kelly used an extension of von Neumann's model (vNM) of measurement to study the dynamical effect of two ``probes" which interact with the system and are designed to ``measure" $x$ and $p$ at the same time.
iii) The role of successive measurements.
A variant of Arthurs-Kelly's model is also studied in Ref. 
\cite{B1}
and interpreted as a sequential measurement of position and momentum.

In the present paper we take up again the 
{\em ``successive-measurement"} form of the problem, 
and employ the vNM as described in 
Refs. \cite{johansen_mello_2008,mello_lasp_aip_2014}, to elaborate on, and investigate further, the UR arising from successive measurements carried out on the {\em same system}.
We may consider this model as extending Arthurs-Kelly's analysis to study two probes interacting with the system at two successive times $t_1$ and $t_2$, so we can exhibit explicitly how the disturbing effect of the first measurement affects the second.

The paper is organized as follows.
In Sec. \ref{succ. meas.} we give an outline of the vNM of measurement.
In the spirit of that model, it is the {\em probe} variables, like $Q_1, Q_2$, that we {\em detect}, in order to uncover information on the system proper:
indeed, the procedure discloses 
the statistical distribution of the eigenvalues of the {\em system} observable 
associated with the second measurement, 
conditioned on the eigenvalues of the {\em system} observable 
for the first measurement to lie within a given resolution
[Eqs. (\ref{p(Q2|Q1=an) strong coupling Q=an}) below]:
the width of this distribution as function of the resolution of the first measurement will constitute the {\em uncertainty relation} of main interest in this paper
[Eqs. (\ref{delta b}) and (\ref{Delta b}) below].
When the second moment of this latter distribution exists, we found the general 
inequality of Eq. (\ref{robertson}) below, that connects the UR with the commutator of the two observables measured successively.
The theory is illustrated in a simplified situation in Sec. \ref{simplified_situation}.
We investigate, in Sec. \ref{[A,B]=0}, the consequence of 
the two observables in question being commuting observables, and then, within a model, we extend the analysis to the general case of an arbitrary commutator
(Sec. \ref{[A,B] arbitr}).
In Sec. \ref{periodic_model} we illustrate 
the formalism 
in the case of the position-like and momentum-like operators defined for Schwinger's model
\cite{schwinger} in a discrete, finite-dimensional Hilbert space.
Within this model, we verify the general 
inequality of Eq. (\ref{robertson}).
In Sec. \ref{application to continuous position-momentum} we apply 
the theory
to the successive measurement of momentum and position in the continuous case.
Finally, we give a summary and our concluding remarks in Sec. \ref{conclu}.
A number of appendices have been included, in order to present some developments without interrupting the main flow of the paper.

\section{von Neumann's Model for the Successive Measurement of two observables
}
\label{succ. meas.}

We briefly describe the successive-measurement model of Refs. \cite{johansen_mello_2008,mello_lasp_aip_2014},
which is an extension to two probes of the vNM of measurement. 
A system $s$ is coupled successively to two 
auxiliary degrees of freedom, or {\em probes}, 
and some properties of the latter are {\em detected} using a 
measuring device:
in other words, {\em it is the probes, not the system proper, which are detected}.
From the detection of the probes 
{\em we can obtain information on the system proper}.
Within the vNM, the combined system 
--system proper plus probes--
is given a dynamical description.

We first define the system observables $\hat{A}, \hat{B}$ as Hermitean operators with the spectral representation
\begin{subequations}
\begin{eqnarray}
\hat{A}= 
\sum_n a_n \mathbb{P}_{a_n} 
\;\;\;\;\;\;
\mathbb{P}_{a_n} = |a_n\rangle \langle a_n|
\; ,
\label{sp repr A} \\
\hat{B}
= \sum_m b_m \mathbb{P}_{b_m} 
\;\;\;\;\;\;
\mathbb{P}_{b_m} = |b_m\rangle \langle b_m|  \; , 
\label{sp repr B}
\end{eqnarray}
\label{sp repr A,B}
\end{subequations}
$a_n$, $b_m$ being the eigenvalues, which, for simplicity, we assume to be 
{\em non-degenerate}, and $\mathbb{P}_{a_n}$, $\mathbb{P}_{b_m}$
the eigenprojectors.

The two probes, assumed to be one-dimensional for simplicity, are
described by the canonical variables 
($\hat{Q}_1, \hat{P}_1$), ($\hat{Q}_2, \hat{P}_2$), respectively.
We consider a model in which their interaction with the system proper is given by
\begin{equation}
\hat{V}
= \epsilon_1 g_1(t) \hat{A}^{\delta a}\hat{P_1} 
+\epsilon_2 g_2(t) \hat{B}\hat{P_2},   \;\;\;\; 0<t_1<t_2,
\label{V 2 probes}
\end{equation}
designed to measure the 
{\em low-resolution observable}
$\hat{A}^{\delta a}$ with the first probe at time $t_1$ and, 
{\em for the same system}, 
the second observable $\hat{B}$ with the second probe at time $t_2$.
The observable $\hat{A}^{\delta a}$ is a low-resolution (resolution $\delta a$)
version of the observable $\hat{A}$, to be described below.
On the other hand, the second observable will be taken as a {\em full resolution}
observable.

The functions $g_1(t)$ and $g_2(t)$ are narrow non-overlapping functions, 
centered around $t=t_1$ and $t=t_2$, respectively, with $0<t_1<t_2$, and 
\begin{equation}
\int_{0}^{\infty} g_i(t') dt' = 1, \;\;\; i=1,2 .
\end{equation}
We neglect the intrinsic dynamics of the various components, so that the interaction
(\ref{V 2 probes}) will be taken as the full Hamiltonian.

The low-resolution observable $\hat{A}^{\delta a}$ is defined in the following way.
We group the eigenvalues $a_n$ into sets of $\delta a +1$ eigenvalues each, centered at $a^{(0)}$, $a^{(1)}$, etc. 
The various intervals are {\em disjoint}.

\begin{figure}[!ht]
\centering
\includegraphics[width=15cm,height=4cm]{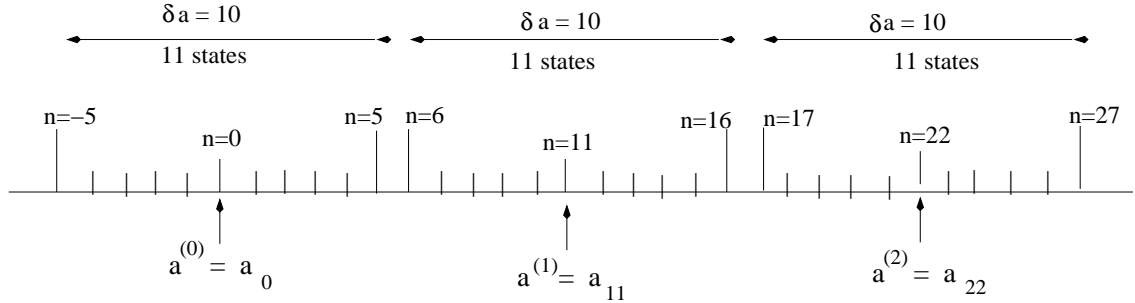}
\caption{Schematic representation of the individual eigenvalues $a_{n'}$
and the intervals centered at $a^{(n)}$ with a width $\delta a = 10$,
used to construct the low-resolution observable defined in the text.}
\label{fig_a_n}
\end{figure}

As illustrated in the example of Fig. \ref{fig_a_n}, two successive interval centers, like 
$a^{(0)}$ and $a^{(1)}$, correspond to $a_n$s whose indices differ by $\delta a +1$.

The meaning of $\delta a$ is the {\em \underline{number of states} around a certain $a^{(n)}$,  different from $a^{(n)}$, that a measurement 
cannot distinguish from $a^{(n)}$};
thus $\delta a$ is the  length of the interval centered at $a^{(n)}$;
$\delta a$ is taken as an {\em even} number, there being $\delta a /2$ states
on each side of $a^{(n)}$, and
$\delta a + 1$ states altogether in the interval around $a^{(n)}$.
The spectral representation of the operators $\hat{A}^{\delta a}$ and $\hat{B}$,
and the eigenprojectors are defined as
\begin{subequations}
\begin{eqnarray}
&& \hat{A}^{\delta a} 
= \sum_{n} a^{(n)}  \mathbb{P}_{a^{(n)}}^{\delta a},  \;\;\;\;\;\;
\mathbb{P}_{a^{(n)}}^{\delta a} 
= \sum_{a_{n'} \in (a^{(n)}, \delta a)} 
\mathbb{P}_{a_{n'}} \; , \;\;\;\;\;\;
\mathbb{P}_{a_{n'}} = |a_{n'}\rangle \langle a_{n'}| , \;\;\; 
\label{sp repr A delta a 0}      \\
&&\hat{B} =\sum_m b_m \mathbb{P}_{b_m}, \hspace{2cm}
\mathbb{P}_{b_m} = |b_m\rangle \langle b_m| \; .
\label{sp repr B 0} 
\end{eqnarray}
\label{sp repr A,B 0}
\end{subequations}
Notice that in the present notation, $a_{n'}$, with a lower index $n'$, designates the $n'$-th eigenvalue of $\hat{A}$, while $a^{(n)}$, with an upper index $(n)$, designates the eigenvalue at the center of the $n$-th interval.

If the $a_n$ level density $\rho(a)$, i.e., the number of states per unit $a$, is approximately constant inside $\delta a$, 
and $D a$ is the extension in $a$ of the $\delta a$ levels,
we may write, approximately
\begin{equation}
\delta a \approx \rho(a) Da \; .
\label{level density}
\end{equation}
While $\delta a$ is dimensionless, $Da$ has the dimensions of the observable $\hat{A}$.
Although in the future we shall use $\delta a$, we might trivially change it to $Da$, according to 
Eq. (\ref{level density}).


The projector $\mathbb{P}_{ a^{(n)}}^{\delta a}$ 
defined  in the second expression of Eq. (\ref{sp repr A delta a 0})
filters coherently the 
$a_{n'}$ components inside the interval $\delta a$ centered at $a^{(n)}$.
The sum in that expression contains
$\delta a +1$ terms; for simplicity, it will often be designated as 
$\sum_{a_{n'} \in (a^{(n)}, \delta a)}$.

The operators $\mathbb{P}_{a^{(n)}}^{\delta a}$ 
have the following properties:

1) They are well defined projector operators, satisfying
\begin{equation}
\mathbb{P}_{a^{(n)}}^{\delta a}
\mathbb{P}_{a^{(n')}}^{\delta a}
= \delta_{n n'}
\mathbb{P}_{a^{(n)}}^{\delta a} \;\;\;\;
\label{P-delta-a-property 1}
\end{equation}

2) They are eigen-projectors of the operator $\hat{A}^{\delta a}$; i.e.,
\begin{equation}
\hat{A}^{\delta a}
\mathbb{P}_{a^{(n)}}^{\delta a}
= a^{(n)}  \mathbb{P}_{a^{(n)}}^{\delta a}
\;\;\;\;\;
\label{P-delta-a-property 2}
\end{equation}

3) However, they are {\em not} eigen-projectors of the operator 
$\hat{A}$; i.e.,
\begin{equation}
\hat{A} \mathbb{P}_{a^{(n)}}^{\delta a}
= \sum_{a_{n'} \in (a^{(n)}, \delta a)} 
a_{n'} \mathbb{P}_{a_{n'}}
\approx a^{(n)}\mathbb{P}_{a^{(n)}}^{\delta a}
\;\;\;\;\;
\label{P-delta-a-property 3}
\end{equation}
the equality sign holding only approximately, if the interval 
$\delta a$  is small enough.

4) They fulfill the completeness relation
\begin{equation}
\sum_{n}\mathbb{P}_{a^{(n)}}^{\delta a}
= \hat{\mathbb{I}} .
\end{equation}

We go back to the state evolution when the Hamiltonian is given by Eq. (\ref{V 2 probes}).
We consider the following initial condition: at $t=0$,
the system is in the state $\rho_s$,
a {\em mixed} state in general, and the two probes, $i=1,2$, are in the {\em pure} Gaussian states
\begin{eqnarray}
\chi_i (Q_i)
= \frac{e^{-\frac{Q_i^2}{4 \sigma_{Q_i}^2}}}{(2 \pi \sigma_{Q_i}^2)^{1/4}}.
\label{chi 1,2}
\end{eqnarray}

Now comes the heart of the procedure.
At time $t_f > t_2$, i.e., after the system-probes interactions are over, we detect, for an {\em individual system} $s$,
a couple of dynamical variables associated with the 
{\em two probes}, which can be done since they commute. 
From the statistical properties of these detected variables 
we uncover information on the statistical properties of the {\em system} variables.

\begin{figure}[h]
\centerline{
\includegraphics[width=15cm,height=5.0cm]
{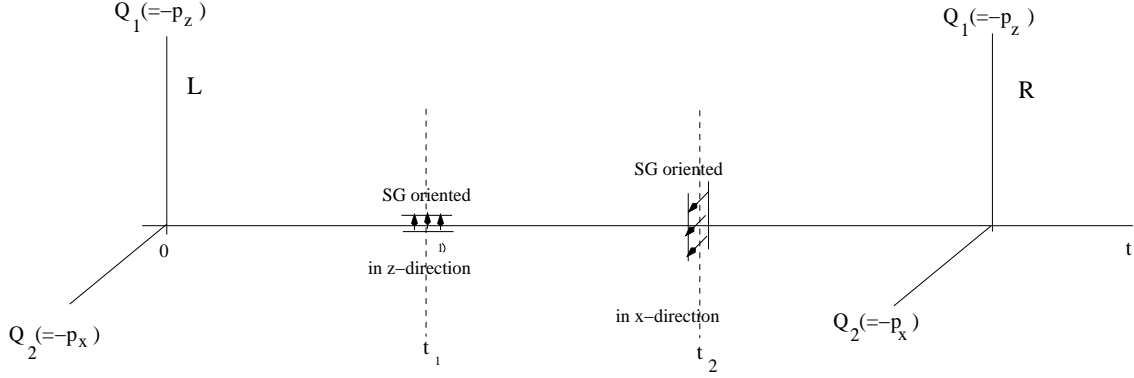}
}
\caption{Illustration of the measurement of 
$\langle \hat{Q_1} \hat{Q_2} \rangle_f$
for an arrangement of two successive SG experiments.
Taken from Ref. \cite{mello_lasp_aip_2014}}.
\label{2SG}
\end{figure}

As an illustration, take the particular 
case $\delta a =0$
and consider the familiar Stern-Gerlach (SG) experiment.
The translation of the general formalism to this problem is through the relations
\begin{subequations}
\begin{eqnarray}
&&\hat{A} \Rightarrow \hat{\sigma_z}, 
\hspace{1.5cm}
\hat{B} \Rightarrow \hat{\sigma_x}  
\label{sigmaz,sigmax}  \\
&&\hat{P_1} \Rightarrow \hat{z},
\hspace{1.5cm}
\hat{P_2} \Rightarrow \hat{x} 
\label{z,x}   \\
&&\hat{Q_1} \Rightarrow -\hat{p_z}, 
\hspace{1cm}
\hat{Q_2} \Rightarrow -\hat{p_x} 
\label{pz,px}
\end{eqnarray}
\end{subequations}
From a measurement of the spatial variables (the ``probes") of 
Eq. (\ref{pz,px}) we can find information on the statistical properties of the 
$z$ and $x$ spin components 
(the ``system proper"), Eqs. (\ref{sigmaz,sigmax}):
we represent schematically, in
Fig. \ref{2SG}, the problem of two successive SG experiments on {\em individual} atoms.

As an example, we explain what we mean by the measurement of 
$\langle Q_1 Q_2 \rangle_f$ ($f$ stands for ``final": see Eq. (\ref{rho f})), 
translating the relevant variables to those associated with 
a SG arrangement. 
Only the axes $Q_1$ and $Q_2$ are shown.
The procedure to measure $\langle \hat{Q_1} \hat{Q_2} \rangle_f$ is:

a) send {\em one} atom from L to R;

b) measure, at R, for {\em that} atom,
$\hat{Q_1} = -\hat{p_z}$ {\em and}
$\hat{Q_2} = -\hat{p_x}$ (recall that $[\hat{Q_1}, \hat{Q_2}] = 0$)
and construct the product $Q_1 Q_2$;

c) repeat the experiment to create an {\em ensemble} of 
${\cal N}$ atoms and construct
\begin{equation}
\langle \hat{Q_1} \hat{Q_2} \rangle_f
= \frac{1}{{\cal N}}\sum_{j=1}^{\cal N} Q_1^j Q_2^j.
\end{equation}
Each $j$ represents a toss of {\em one atom}.
Similarly, we can construct the conditioned expectation value
\begin{equation}
E(Q_2|Q_1)=E(-p_x|-p_z)
=\frac{1}{{\cal N}}\sum_{j=1}^{{\cal N}}(-p_x)^{j}|_{p_z}  \; .
\label{E(Q1|Q2)SG}
\end{equation}


Having illustrated the basic idea in a familiar setup,
we return to the general formalism.
We study the probability distribution (pd) 
of the second probe position
$Q_2$, {\em conditioned} on the first probe position taking the value $Q_1$.         
In App. \ref{derivation p(Q2|Q1)} we show that the result is
\begin{equation}
p(Q_2|Q_1)
= \frac{ 
\sum_{m, n, n'} g_{nn'}(\epsilon_1 / \sigma_{Q_1}) \;
{\rm Tr}_s(\rho_s 
\mathbb{P}_{a^{(n')}}^{\delta a} 
\mathbb{P}_{b_m} 
\mathbb{P}_{a^{(n)}}^{\delta a}) \;
\frac{{\rm e}^{-\frac{\left(Q_1 - \epsilon_1\frac{a^{(n)}
+a^{(n')}}{2} \right)^2}{2 \sigma_{Q_1}^2}}}{\sqrt{2 \pi \sigma_{Q_1}^2}} \;
\frac{{\rm e}^{-\frac{\left(Q_2 - \epsilon_2 b_m \right)^2}{2  \sigma_{Q_2}^2}}}{\sqrt{2 \pi \sigma_{Q_2}^2}}
}
{ 
\sum_n {\rm Tr}_s (\rho_s \mathbb{P}_{a^{(n)}}^{\delta a}) \;
\frac{{\rm e}^{-\frac{\left(Q_1 - \epsilon_1 a^{(n)}\right)^2}{2 \sigma_{Q_1}^2}}}{\sqrt{2 \pi \sigma_{Q_1}^2}}
} \; .
\label{p(Q2|Q1)}
\end{equation}
Here, the factor  $g_{n, n'}(\epsilon_1 / \sigma_{Q_1})$ 
is given by
\begin{equation}
g_{n, n'}(\epsilon_1 / \sigma_{Q_1})
= {\rm e}^{-\frac{\epsilon_1^2}{2\sigma_{Q_1}^2}(a^{(n)} - a^{(n')})^2} \;.
\label{g}
\end{equation}

In the limit in which the first probe is {\em weakly coupled} to the system,
$\epsilon_1 / \sigma_{Q_1} \ll 1$, we have
$g_{nn'}(\epsilon_1 / \sigma_{Q_1}) \approx 1$,
and Eq. (\ref{p(Q2|Q1)}) reduces to
\begin{subequations}
\begin{eqnarray}
p(Q_2|Q_1) &\approx &
\frac{ 
\sum_{m,n, n'} 
{\rm Tr}_s\left[(         
\rho_s 
\mathbb{P}_{a^{(n')}}^{\delta a}           
\mathbb{P}_{b_m}
\mathbb{P}_{a^{(n)}}^{\delta a}  \right]  \;
\frac{{\rm e}^{-\frac{Q_1^2}{2 \sigma_{Q_1}^2}}}{\sqrt{2 \pi \sigma_{Q_1}^2}} \;
\frac{{\rm e}^{-\frac{(Q_2 - \epsilon_2 b_m)^2}{2 \sigma_{Q_2}^2}}}{\sqrt{2 \pi \sigma_{Q_2}^2}} \;
}
{ 
\sum_{n} {\rm Tr}_s (\rho_s 
\mathbb{P}_{a^{(n)}}^{\delta a})         \;
\frac{{\rm e}^{-\frac{Q_1^2}{2 \sigma_{Q_1}^2}}}{\sqrt{2 \pi \sigma_{Q_1}^2}}
}    
\label{p(Q2|Q1) weak coupling a}  \\
&=&
\frac{ 
\sum_{m} 
{\rm Tr}_s\left(        
\rho_s            
\mathbb{P}_{b_m}
\right)  \;
\frac{{\rm e}^{-\frac{(Q_2 - \epsilon_2 b_m)^2}{2 \sigma_{Q_2}^2}}}{\sqrt{2 \pi \sigma_{Q_2}^2}} \;
}
{ 
\sum_{n} {\rm Tr}_s (\rho_s 
\mathbb{P}_{a^{(n)}}^{\delta a})  \;
}   
\label{p(Q2|Q1) weak coupling b}     \\
&=& 
\sum_{m} W_{b_m} \cdot |\chi(Q_2  - \epsilon_2 b_m))|^2 \; .
\label{p(Q2|Q1) weak coupling c}
\end{eqnarray}
\label{p(Q2|Q1) weak coupling}
\end{subequations}
Here, $W_{b_m}$ is the Born probability for the result $b_m$
in the original system state $\hat{\rho}_s$ and
$|\chi(Q_2  - \epsilon_2 b_m))|^2$
is the original $Q_2$ probability density displaced by the amount $\epsilon_2 b_m$
(its width is $\sigma_{Q_2}$);
the result is {\em insensitive to the presence of the first probe}, as it has to be.
A word of caution is in order.
In going from Eq. (\ref{p(Q2|Q1)}) to 
Eq. (\ref{p(Q2|Q1) weak coupling c}) we freely interchanged the order of limits:
   i) number of terms $\to \infty$,    
   ii) $\epsilon_1 / \sigma_{Q_1} \to 0$.   
We remark that, according to Ref. \cite{apostol}, a sufficient condition for the validity of such an interchange is that 
the series be ``uniformly convergent".

In the opposite limit
--the limit of interest in the present paper, which will be assumed henceforth--
in which the first probe is {\em strongly coupled} to the system,
$\epsilon_1 / \sigma_{Q_1} \gg 1$, we have 
$g_{n n'}(\epsilon_1 / \sigma_{Q_1}) \approx \delta_{n n'}$,
and Eq. (\ref{p(Q2|Q1)}) reduces to
\begin{eqnarray}
p(Q_2|Q_1)
\approx
\frac{ 
\sum_{m,n} 
{\rm Tr}_s 
\left( \rho_s 
\mathbb{P}_{a^{(n)}}^{\delta a}           
\mathbb{P}_{b_m} 
\mathbb{P}_{a^{(n)}}^{\delta a}
\right)  \;
\frac{{\rm e}^{-\frac{\left(Q_1 - \epsilon_1 a^{(n)} \right)^2}{2 \sigma_{Q_1}^2}}}{\sqrt{2 \pi \sigma_{Q_1}^2}} \;
\frac{{\rm e}^{-\frac{\left(Q_2 - \epsilon_2 b_{m} \right)^2}{2 \sigma_{Q_2}^2}}}{\sqrt{2 \pi \sigma_{Q_2}^2}}
}
{ 
\sum_{n} {\rm Tr}_s (\rho_s 
\mathbb{P}_{a^{(n)}}^{\delta a}            
) \;
\frac{{\rm e}^{-\frac{\left(Q_1 - \epsilon_1 a^{(n)}\right)^2}{2 \sigma_{Q_1}^2}}}{\sqrt{2 \pi \sigma_{Q_1}^2}}
} \; .
\label{p(Q2|Q1) strong coupling}
\end{eqnarray}
In this strong-coupling limit, the Gaussians in the variable $Q_1$ appearing in the above equation are widely separated from one another. If we take, for instance, $Q_1 = \epsilon_1 a^{(n_0)}$, only the term $n=n_0$ survives in both the numerator and denominator of 
Eq. (\ref{p(Q2|Q1) strong coupling}), and
$p(Q_2|Q_1 = \epsilon_1 a^{(n_0)})$
can be given the various equivalent forms that follow
\begin{subequations}
\begin{eqnarray}
p(Q_2|Q_1 = \epsilon_1 a^{(n_0)})
&\approx& \sum_m
\frac{  
{\rm Tr}_s\left[
\left(\mathbb{P}_{a^{(n_0)}}^{\delta a}         
\rho_s 
\mathbb{P}_{a^{(n_0)}}^{\delta a}\right) \mathbb{P}_{b_m} \right] 
}
{{\rm Tr}_s (
\rho_s 
\mathbb{P}_{a^{(n_0)}}^{\delta a}) \;
}   
\; \cdot 
\frac{{\rm e}^{-\frac{\left(Q_2 - \epsilon_2 b_{m} \right)^2}
{2 \sigma_{Q_2}^2}}}{\sqrt{2 \pi \sigma_{Q_2}^2}} 
\label{p(Q2|Q1) strong coupling Q=an a} \\
&\equiv& \sum_m {\rm Tr}_s\left(\rho_s^{a^{(n_0)},\delta a} \; \mathbb{P}_{b_m} \right)
\cdot 
\frac{{\rm e}^{-\frac{\left(Q_2 - \epsilon_2 b_{m} \right)^2}
{2 \sigma_{Q_2}^2}}}{\sqrt{2 \pi \sigma_{Q_2}^2}}
\label{p(Q2|Q1) strong coupling Q=an b} \\
&=& {\rm convolution \; of \;the \; original} \; 
Q_2 \; {\rm pd \; and \; the} \;
\nonumber \\
&& \;\;\;\;\;\;  b_m  \;{\rm pd \; in \; the \; perturbed \; system \; state} \; \rho_s^{a^{(n_0)}, \delta a}  
\label{p(Q2|Q1) strong coupling Q=an c}   \\
&=&
\sum_m \frac{  
{\rm Tr}_s
\left( \rho_s 
\mathbb{P}_{a^{(n_0)}}^{\delta a}            
           \mathbb{P}_{b_{m}} 
\mathbb{P}_{a^{(n_0)}}^{\delta a}            
\right)
}
{
{\rm Tr}_s 
\left(\rho_s 
 \mathbb{P}_{a^{(n_0)}}^{\delta a}              
\right)
}  
\; \frac{{\rm e}^{-\frac{\left(Q_2 - \epsilon_2 b_{m} \right)^2}
{2 \sigma_{Q_2}^2}}}{\sqrt{2 \pi \sigma_{Q_2}^2}} 
\label{p(Q2|Q1) strong coupling Q=an d}    \\
&=&
\sum_m \frac{  
{\cal W}(b_m; a^{(n_0)}, \delta a)
}
{
{\rm Tr}_s 
\left(\rho_s 
 \mathbb{P}_{a^{(n_0)}}^{\delta a}              
\right)
}  
\;\frac{{\rm e}^{-\frac{\left(Q_2 - \epsilon_2 b_{m} \right)^2}
{2 \sigma_{Q_2}^2}}}{\sqrt{2 \pi \sigma_{Q_2}^2}}
\label{p(Q2|Q1) strong coupling Q=an e} \\
&=&
\sum_m 
{\cal W}(b_m | a^{(n_0)}, \delta a)
\; \frac{{\rm e}^{-\frac{\left(Q_2 - \epsilon_2 b_{m} \right)^2}
{2 \sigma_{Q_2}^2}}}{\sqrt{2 \pi \sigma_{Q_2}^2}}
\label{p(Q2|Q1) strong coupling Q=an f}  \\
&=& 
{\rm convolution \; of \; the \; original} \; 
Q_2 \; {\rm pd \; and \; the} \; 
\nonumber \\
&& \;\;\;\;\;\; {\rm distribution} 
\; {\cal W}(b_m |  a^{(n_0)}, \delta a) \; .
\label{p(Q2|Q1) strong coupling Q=an g}
\end{eqnarray}
\label{p(Q2|Q1=an) strong coupling Q=an}
\end{subequations}
In this strong-coupling limit, the probability distribution of $Q_2$, conditioned on $Q_1= \epsilon_1 a^{(n_0)}$,
is expressed, in
Eqs. 
(\ref{p(Q2|Q1) strong coupling Q=an d}) and 
(\ref{p(Q2|Q1) strong coupling Q=an e}),
in terms of what is known as ``Wigner's formula" [\onlinecite{wigner63}],
\begin{equation}
{\cal W}(b_m; a^{(n_0)}, \delta a)
={\rm Tr}_s\left( \rho_s 
\mathbb{P}_{a^{(n_0)}}^{\delta a}      
\mathbb{P}_{b_{m}} 
\mathbb{P}_{a^{(n_0)}}^{\delta a}      \right) \; ,
\label{wigner-formula}
\end{equation}
for the joint probability of finding {\em first} $a_n$ in the interval $\delta a$
around $a^{(n_0)}$
in an experiment with {\em resolution}
$\delta a$,
and {\em then} $b_m$.
Eq. (\ref{wigner-formula}) is actually a generalization of Wigner's formula:
it reduces to the standard one when $\delta a =0$.
Wigner's formula, obtained by Wigner using the {\em collapse} postulate when the measuring probes are {\em not} included, 
appears here as a {\em property of the probes}
[the LHS of Eqs. (\ref{p(Q2|Q1=an) strong coupling Q=an})], 
{\em no collapse} having ever been
assumed.
In the language of Wigner's formula, the perturbed system state after the first measurement
(a selective projective measurement) is seen,
from Eqs. (\ref{p(Q2|Q1) strong coupling Q=an a}) and
(\ref{p(Q2|Q1) strong coupling Q=an b})
to be given by 
\begin{equation}
\rho_s^{a^{(n_0)},\delta a}
= \frac{  
\mathbb{P}_{a^{(n_0)}}^{\delta a}                    
\; \rho_s \;
\mathbb{P}_{a^{(n_0)}}^{\delta a}                            
}
{{\rm Tr}_s (\rho_s 
\mathbb{P}_{a^{(n_0)}}^{\delta a}            
) 
} \; :
\label{perturbed rho}
\end{equation}
this is the ``state projection", or ``collapse" postulate referred to above.

In Eqs. (\ref{p(Q2|Q1) strong coupling Q=an a}), 
(\ref{p(Q2|Q1) strong coupling Q=an d}), and
(\ref{p(Q2|Q1) strong coupling Q=an e}),
${\rm Tr}_s\left(\rho_s 
 \mathbb{P}_{a^{(n_0)}}^{\delta a} \right)$
is the Born probability for the result $a_n \in (a^{(n_0)}, \delta a)$ in the original state $\rho_s$.

In Eq. (\ref{p(Q2|Q1) strong coupling Q=an f}),
${\cal W}(b_m |  a^{(n_0)}, \delta a)$
is then the probability, given by ``Wigner's formula", of finding $b_m$ {\em conditioned} by having found $a_n \in (a^{(n_0)}, \delta a)$, i.e.,
\begin{equation}
{\cal W}(b_m |a^{(n_0)}, \delta a)
\equiv \frac{  
{\rm Tr}_s\left( \rho_s 
\mathbb{P}_{a^{(n_0)}}^{\delta a}                
\mathbb{P}_{b_{m}} 
\mathbb{P}_{a^{(n_0)}}^{\delta a}                 \right)
}
{{\rm Tr}_s \left(\rho_s 
\mathbb{P}_{a^{(n_0)}}^{\delta a} 
\right)
}  
\; .
\label{cond_wigner_formula}
\end{equation}

In Eq. (\ref{p(Q2|Q1) strong coupling Q=an b}),
$p(Q_2|Q_1 = \epsilon_1 a^{(n_0)})$
is expressed as the 
convolution of the original $Q_2$ pd and the
$b_m$ pd in the perturbed system state $\rho_s^{a^{(n_0)}, \delta a}$ .
Similarly, in Eq. (\ref{p(Q2|Q1) strong coupling Q=an f}),
$p(Q_2|Q_1 = \epsilon_1 a^{(n_0)})$
is expressed as the 
convolution of the original $Q_2$ pd and the distribution 
${\cal W}(b_m |  a^{(n_0)}, \delta a)$.

The conditioned Wigner formula of Eq. (\ref{cond_wigner_formula}) 
is a probability distribution for the variable $b_m$, and is a function of 
the resolution $\delta a$
of the first measurement.
The width of this probability distribution, as function of $\delta a$, constitutes an 
{\em uncertainty relation}.
As a measure of this width, we shall speak of the quantity 
$\delta b$ as 
\begin{eqnarray}
&& \delta b = number \; of \; states \;  b_m , 
{\rm as \; a \; function \; of \;} \delta a,  {\rm\; over \;  which} \; 
\nonumber  \\
&& \hspace{1cm} {\cal W}(b_m |a^{(n_0)}, \delta a)\;  
{\rm is \; appreciably \; different \; from \; zero}.
\label{delta b}
\end{eqnarray}
We may also consider, as an alternative measure of this width, the standard deviation for the probability distribution 
${\cal W}(b_m |a^{(n_0)}, \delta a)$, defined, when it exists, as
\begin{eqnarray}
\Delta B \equiv \sqrt{{\rm var}(b_m)s \; {\rm for} \; 
{\cal W}(b_m |a^{(n_0)}, \delta a)}
\label{Delta b}
\end{eqnarray}
as in Eqs. (\ref{var(Q2|Q1) strong coupling Q=an e}),
(\ref{var(Q2|Q1) strong coupling Q=an f}) and
(\ref{varBpert=F(da)})
below.
Notice that, while $\Delta B$ depends on the {\em actual values of the spectral quantities} $b_m$, $\delta b$ does not, since it is the {\em number of states} in the interval defined in Eq. (\ref{delta b}).

Specific examples of this relation are implemented in 
the following sections.

The message of Eqs. (\ref{p(Q2|Q1=an) strong coupling Q=an})
is that from the LHS, $p(Q_2|Q_1 = \epsilon_1 a^{(n_0)})$, measured for the {\em probes}, we can extract 
${\cal W}(b_m | a^{(n_0)}, \delta a)$ for the {\em system}.
To see this, we multiply both sides of 
Eq. (\ref{p(Q2|Q1) strong coupling Q=an f}) by ${\rm e}^{iK_2Q_2}$ 
and integrate over $Q_2$, to find 
\begin{eqnarray}
\int_{-\infty}^{\infty} {\rm e}^{iK_2Q_2}
p(Q_2|Q_1=\epsilon_1 a^{(n_0)}) dQ_2
&=&
\sum_{m} {\cal W}(b_m|a^{(n_0)}, \delta a) \;
\int_{-\infty}^{\infty}
\frac{{\rm e}^{-\frac{\left(Q_2 - \epsilon_2 b_m \right)^2}{2  \sigma_{Q_2}^2}}}
{\sqrt{2 \pi \sigma_{Q_2}^2}}
{\rm e}^{iK_2Q_2} dQ_2
\nonumber \\
&=& \left[ \sum_{m} {\cal W}(b_m|a^{(n_0)}, \delta a) e^{iK_2 \epsilon_2 b_m}\right]
e^{-\frac12 K_2^2 \sigma_{Q_2}^2}
\label{ch function 1}
\end{eqnarray}
The LHS 
of Eq. (\ref{ch function 1}) is the ``characteristic function" $\widetilde{p}(K_2|Q_1=\epsilon_1 a^{(n_0)})$.
The square bracket on the RHS  is the ``characteristic function"
$\widetilde{{\cal W}}(\epsilon_2 K_2 |a^{(n_0)}, \delta a)$.
We then have
\begin{eqnarray}
\widetilde{p}(K_2|Q_1=\epsilon_1 a^{(n_0)})
= \widetilde{{\cal W}}(\epsilon_2 K_2 |a^{(n_0)}, \delta a)
\;
e^{-\frac12 K_2^2 \sigma_{Q_2}^2}\; ,
\label{ch function 2}
\end{eqnarray}
a result which could have been anticipated from the convolution theorem.
We have thus found that, in the strong-coupling limit, the characteristic function of the {\em probe} variable $Q_2$ gives directly the characteristic function of the {\em system} variable $b_m$.


More specifically, from 
Eqs. (\ref{p(Q2|Q1=an) strong coupling Q=an}) for the strong-coupling limit we obtain, for instance, if the various moments are well defined,
\begin{subequations}
\begin{eqnarray}
\frac{E(Q_2|Q_1 = \epsilon_1 a^{(n_0)})}{\epsilon_2}
&=&
{\rm Tr}_s \left(\rho_s^{a^{(n_0)},\delta a}\hat{B}\right)  
\label{E(Q2|Q1) strong coupling Q=an a} \\
&=&
\sum_m 
{\cal W}(b_m | a^{(n_0)}, \delta a)
\; b_m
\label{E(Q2|Q1) strong coupling Q=an b}  \\
&=& 
\Big\{ {\rm 1st \; moment \; of \; the \;} b_m {\rm s \; for \; the \; pd} 
\; {\cal W}(b_m | a^{(n_0)}, \delta a) \Big\}
\nonumber  \\
\frac{E(Q_2^2|Q_1 = \epsilon_1 a^{(n_0)})}{\epsilon_2^2}
&=&
{\rm Tr}_s \left( \rho_s^{a^{(n_0)},\delta a} \hat{B}^2 \right) 
+  \left(\frac{\sigma_{Q_2}}{\epsilon_2}\right)^2
\label{E(Q2|Q1) strong coupling Q=an c}  \\
&=&
\sum_m 
{\cal W}(b_m | a^{(n_0)}, \delta a)
\; b_m^2
+ \left(\frac{\sigma_{Q_2}}{\epsilon_2}\right)^2
\label{E(Q2^2|Q1) strong coupling Q=an d}  \\
&=& 
\Big\{ {\rm 2nd \; moment \; of \; the \;} b_m {\rm s \; for \; the \; pd} 
\; {\cal W}(b_m | a^{(n_0)}, \delta a) \Big\}
+\left(\frac{\sigma_{Q_2}}{\epsilon_2}\right)^2 \; ,
\nonumber \\
\frac{{\rm var}(Q_2|Q_1 = \epsilon_1 a^{(n_0)})}{\epsilon_2^2}
&=& ({\rm var}\hat{B})_{\rho_s^{a^{(n_0)},\delta a}}
+ \left(\frac{\sigma_{Q_2}}{\epsilon_2}\right)^2
\equiv F(\delta a) + \left(\frac{\sigma_{Q_2}}{\epsilon_2}\right)^2
\label{var(Q2|Q1) strong coupling Q=an e} \\
&=&
\sum_m {\cal W}(b_m |a^{(n_0)},\delta a) b_m^2 
- \left[\sum_m {\cal W}(b_m |a^{(n_0)},\delta a) b_m \right]^2
+ \left(\frac{\sigma_{Q_2}}{\epsilon_2}\right)^2    
\label{var(Q2|Q1) strong coupling Q=an f}  \\
&=& \Big\{ {\rm variance \; of \; the \;} b_m {\rm s \; for \; the \; pd} 
\; {\cal W}(b_m |a^{(n_0)},\delta a) \Big\}
+\left(\frac{\sigma_{Q_2}}{\epsilon_2}\right)^2 \; .
\nonumber
\end{eqnarray}
\label{E(Q2^2|Q1=an) strong coupling  Q=an}
\end{subequations}

The message of Eqs.
(\ref{E(Q2^2|Q1=an) strong coupling  Q=an})
is that the first and second
moments and the variance of the {\em probe} position $Q_2$,
conditioned on $Q_1 = \epsilon_1 a^{(n_0)}$,
which is what we {\em detect}
within the spirit of the vNM of measurement,
give information, in the strong-coupling limit, on the first and second moments and the variance, respectively,
of the {\em system} variables $b_m$s, distributed according to
${\cal W}(b_m | a^{(n_0)}, \delta a)$.

The variance of the $b_m$s as function of $\delta a$ 
is the {\em uncertainty relation} mentioned in Eq. (\ref{Delta b}): 
it gives the resulting ${\rm var} \hat{B}$
in the perturbed state $\rho_s^{a^{(n_0)}, \delta a}$
in terms of the resolution $\delta a$ 
of the first measurement
[Eq. (\ref{var(Q2|Q1) strong coupling Q=an e})].
This relation is denoted by the function $F(\delta a)$, i.e.,
\begin{equation}
({\rm var} \hat{B})_{\rho_s^{a^{(n_0)}, \delta a}}
=F({\delta a})
\; .
\label{varBpert=F(da)}
\end{equation}
This function can be found operationally, for every $\delta a$, by means of an experiment consisting of a large sample of
{\em ${\cal N}$ ``tosses" of an individual system $s$}, 
measuring, over this sample, the quantity 
${\rm var}(Q_2|Q_1 = \epsilon_1 a^{n_0})$
appearing on the LHS of 
Eq. (\ref{var(Q2|Q1) strong coupling Q=an e}),
as illustrated in Fig. \ref{2SG} for a SG arrangement.
The whole experiment can be repeated for various values of $\delta a$, and the function  
$F(\delta a)$ can then be constructed.

We conclude this section by noting that,
using Robertson's inequality \cite{robertson}, we can write a general 
{\em connection between the UR}
$({\rm var} \hat{B})_{\rho_s^{a^{(n_0)}, \delta a}}$
of Eq. (\ref{varBpert=F(da)}), when the variance exists, and the {\em commutator} 
$[\hat{A}, \hat{B}]$
of the two observables measured successively, as
\begin{equation}
({\rm var} \hat{B})_{\rho_s^{a^{(n_0)}, \delta a}}
\ge \frac14 \frac
{ \left| \left< \left[  \hat{A}, \hat{B}  \right]  \right>_{   \rho_s^{a^{(n_0)}, \delta a}} \right|^2}
{\left( {\rm var} \hat{A}  \right)_{   \rho_s^{a^{(n_0)}, \delta a}    }} \; .
\label{robertson}
\end{equation}
This inequality is verified in Sec. \ref{periodic_model}, where we study a periodic model in a finite-dimensional Hilbert space.

A comment on the above equation is in order.
If, in Eq. (\ref{robertson}), we had taken, instead of $\hat{A}$, 
$\hat{A}^{\delta a}$, on the basis that 
$\hat{A}^{\delta a}$ is the observable ``actually measured", 
then, passing the denominator to the LHS we would have found $0\ge 0$, which is trivially true.
In contrast, using $\hat{A}$ we get a non trivial result.

On the other hand, we emphasize that the LHS of Eq. (\ref{robertson})
is certainly the quantity we want, as it relates
$({\rm var} \hat{B})_{\rho_s^{a^{(n_0)}, \delta a}}$ 
for the second observable with the resolution 
$\delta a$ of the first measurement.

The assertion made two paragraphs above is proved as follows.
From Eq. (\ref{P-delta-a-property 2}) we have
\begin{eqnarray}
\hat{A}^{\delta a}
\mathbb{P}_{a^{(n)}}^{\delta a}
= \mathbb{P}_{a^{(n)}}^{\delta a} \hat{A}^{\delta a}
=a^{(n)}  \mathbb{P}_{a^{(n)}}^{\delta a}
\end{eqnarray}
so  that, from Eq. (\ref{perturbed rho})
\begin{eqnarray}
\hat{A}^{\delta a}\hat{\rho}^{a^{(n)},\delta a}
= \hat{\rho}^{a^{(n)},\delta a} \hat{A}^{\delta a}
= a^{(n)}  \hat{\rho}^{a^{(n)},\delta a}
\end{eqnarray}
and hence
\begin{subequations}
\begin{eqnarray}
\langle  (\hat{A}^{\delta a})^r  \rangle_{\hat{\rho}^{a^{(n)},\delta a}}
&=& (a^{(n)})^r \;\;\; \Longrightarrow \;\;\; ({\rm var}\hat{A}^{\delta a})
_{\hat{\rho}^{a^{(n)},\delta a}} = 0. \\
\langle  \hat{A}^{\delta a} \hat{B} -  \hat{B}\hat{A}^{\delta a}  \rangle
_{\hat{\rho}^{a^{(n)},\delta a}} 
&=& 0
\end{eqnarray}
\end{subequations}

\section{Wigner's Formula in a simplified situation}
\label{simplified_situation}

As a first simplification to illustrate the above formalism,
let the original system state be the {\em pure state} 
$\rho_s =|\psi \rangle_s \; _s\langle \psi |$.
As mentioned above, we assume strong coupling between the system and the first probe, as we shall always do in the present paper.
In the context of Wigner's formula, the perturbed
system state after the first measurement
(a selective projective measurement)
is given by Eq. (\ref{perturbed rho}) as
\begin{subequations}
\begin{eqnarray}
|\psi^{a^{(n_0)}, \delta a} \rangle_s
&=& \frac{\mathbb{P}_{a^{(n_0)}}^{\delta a} |\psi \rangle_s}
{_s\langle \psi | \mathbb{P}_{a^{(n_0)}}^{\delta a}
 |\psi \rangle_s^{1/2}} \\
&=& 
\frac{\sum_{a_n \in (a^{(n_0)}, \delta a)} 
\mathbb{P}_{a_n} |\psi \rangle_s }  
{\left[ \sum_{{a_{n'}}\in (a^{(n_0)}, \delta a)} 
\; _s\langle \psi | \mathbb{P}_{a_{n'}} |\psi \rangle_s
\right]^{1/2}} \\
&=& 
\frac{\sum_{a_n \in (a^{(n_0)}, \delta a)}
| a_{n} \rangle \langle a_{n} |\psi \rangle_s }  
{\Big[ \sum_{{a_{n'}}\in (a^{(n_0)}, \delta a)} 
|\langle a_{n'} |\psi \rangle_s|^2
\Big]^{1/2}} \; .
\end{eqnarray}
\label{pert psi non-deg proj Da neq 0}
\end{subequations}
The results of Eqs. (\ref{pert psi non-deg proj Da neq 0}) 
apply regardless of the observable that is measured  next.

The conditioned Wigner formula, 
Eq. (\ref{cond_wigner_formula}), then gives
\begin{subequations}
\begin{eqnarray}
{\cal W}(b_m | a^{(n_0)}, \delta a)
&=& \;    _s\langle \psi^{{a^{(n_0)}, \delta a}}  |
 \mathbb{P}_{b_m}  
|\psi^{ a^{(n_0)}, \delta a }  \rangle_s \\
&=& 
\frac{\sum_{a_n, a_{n'} \in (a^{(n_0)}, \delta a)} \;
_s\langle \psi|a_{n'} \rangle
\langle a_{n'} |b_m \rangle
\langle b_m|a_n \rangle
\langle a_n| \psi\rangle_s
}
{\sum_{{a_{n''}}\in (a^{(n_0)}, \delta a)} 
|\langle a_{n''} |\psi \rangle_s|^2}   \\
&\equiv& \frac{N(b_m)}{D} ,
\end{eqnarray}
\label{W(b|da)}
\end{subequations}
where
\begin{subequations}
\begin{eqnarray}
N(b_m) &=&  \Big| 
\sum_{{a_{n}}\in (a^{(n_0)}, \delta a)}  
\langle b_m|a_n \rangle
\langle a_n| \psi\rangle_s \Big|^2        
\label{N(bm)}    \\
D &=& \sum_{{a_{n}}\in (a^{(n_0)}, \delta a)} 
|\langle a_{n} |\psi \rangle_s|^2 \; .
\label{D}
\end{eqnarray}
\label{N(bm),D}
\end{subequations}
As mentioned in relation with Eqs. (\ref{sp repr A,B}), the eigenvalues $a_n$, $b_m$
are assumed to be non degenerate.

Just as in the previous section, $\delta a$ denotes the number of states
around $a^{(n_0)}$, different from $a^{(n_0)}$, that the measurement cannot distinguish from $a^{(n_0)}$: it serves to define the low-resolution observable 
$\hat{A}^{\delta a}$,
Eq. (\ref{sp repr A delta a 0})
which is measured first; 
it is not related to the properties of the original system state 
$| \psi \rangle_s$ .
Regarding the original system state $| \psi \rangle_s$ , we assume,
as a second simplification, the particular case in which its components $\langle a_n| \psi\rangle_s$ 
are {\em real and positive} and {\em their distribution is
centered precisely at $a^{(n_0)}$}, with a width that we designate as $\sigma_a$.

\begin{figure}[!ht]
\centering
\includegraphics[width=15cm,height=5cm]{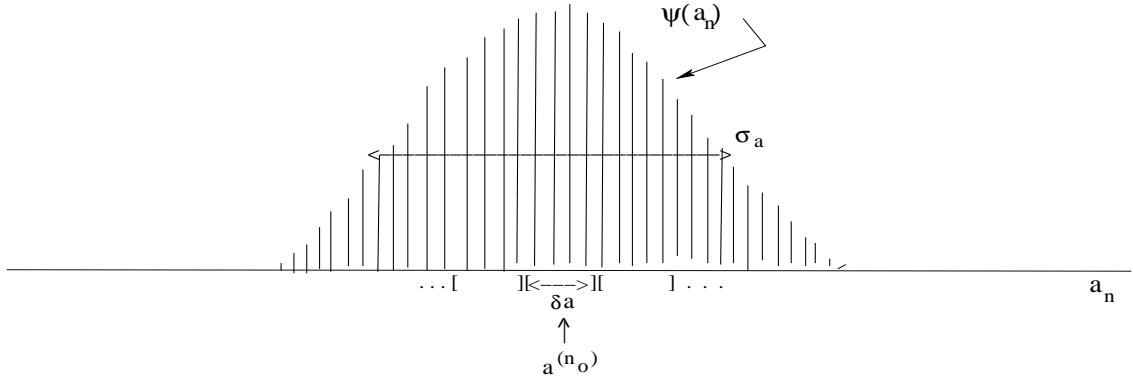}
\caption{Schematic representation of the components 
$\langle a_n | \psi \rangle_s \equiv \psi_s(a_n)$ 
of the system original state, 
in the case $\delta a \ll \sigma_a$.
I.e., the system wavefunction $\psi_s(a_n)$ has a {\em large spread} 
$\sigma_a$ compared
with the resolution $\delta a$ of the first measurement.
The wavefunction $\psi_s(a_n)$ is assumed centered at the same value 
$a^{(n_0)}$ around which the first low-resolution measurement is performed.
}
\label{fig1}
\end{figure}

We consider two extreme situations:

i) $\delta a \ll \sigma_a$ and 
$\langle a_n | \psi \rangle_s$ is almost constant inside $\delta a$;
this situation is illustrated schematically in Fig. \ref{fig1}.

\begin{figure}[!ht]
\centering
\includegraphics[width=16cm,height=7cm]{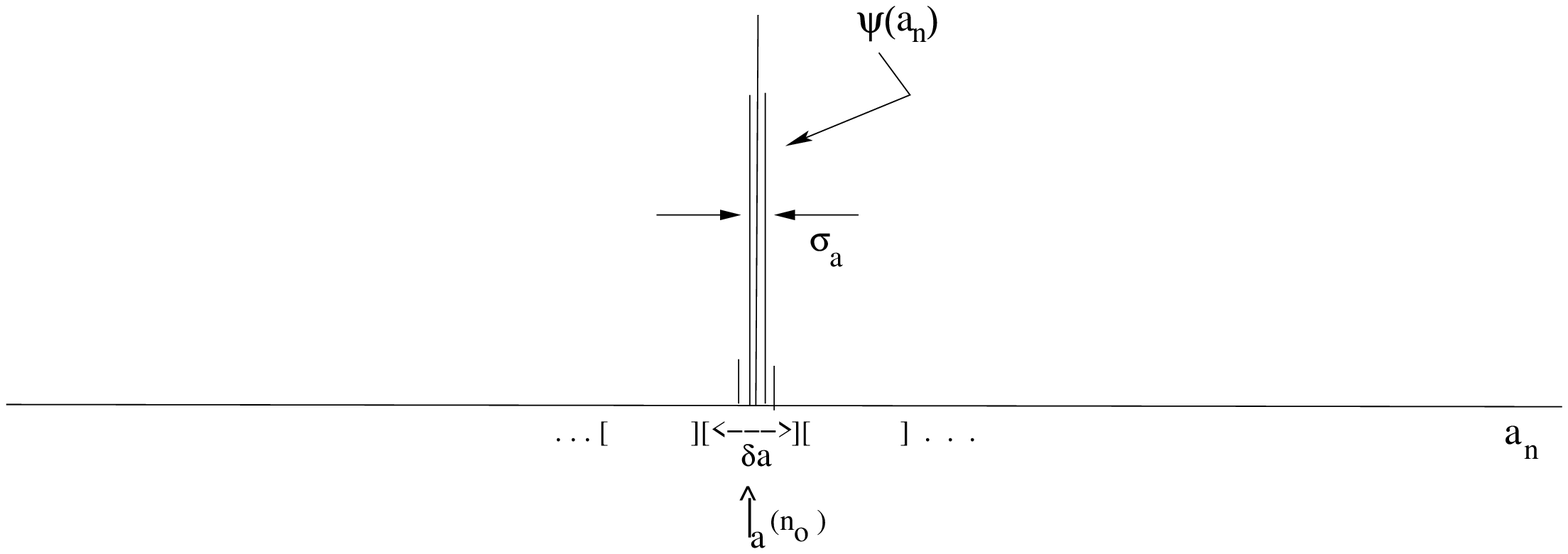}
\caption{Schematic representation of the components 
$\langle a_n | \psi \rangle_s \equiv \psi_s(a_n)$ 
of the system original state, 
in the case $\delta a \gg \sigma_a$.
I.e., the system wavefunction $\psi_s(a_n)$ has a {\em small spread} 
$\sigma_a$ compared
with the resolution $\delta a$ of the first measurement.
The wavefunction $\psi_s(a_n)$ is assumed centered at the same value 
$a^{(n_0)}$ around which the first low-resolution measurement is performed.}
\label{fig2}
\end{figure}

2) $\delta a \gg \sigma_a$, so that the distribution of $\langle a_n | \psi \rangle_s$ is completely contained inside $\delta a$,
as illustrated in Fig. \ref{fig2}.

The quantities $N(b_m)$ and $D$ of Eq. (\ref{N(bm),D}) take the form
\begin{subequations}
\begin{eqnarray}
N(b_m) 
&\approx& \left\{
\begin{array}{cccc} 
\big| \psi(a^{(n_0)}) \big|^2 \;
\Big| 
\sum_{{a_{n}}\in (a^{(n_0)}, \delta a)}  
\langle b_m|a_n \rangle \Big|^2 & {\rm for} & \delta a \ll  \sigma_a, &      \\
{\rm provided} & \langle a_n | \psi \rangle_s &
{\rm almost \; const} & {\rm inside} \; \delta a    
\label{N a}   \\
\Big| 
\sum_{{\rm all} \; {a_{n}}}  
\langle b_m|a_n \rangle
\langle a_n| \psi\rangle_s \Big|^2
 = |\psi(b_m)|^2 \; , & {\rm for} &
\delta a \gg \sigma_a \; .  & 
\label{N b}  \\
\end{array}
\right.
\label{N}   \\ 
\nonumber \\
D &\approx& 
\left\{
\begin{array}{cccc}
\big| \psi(a^{(n_0)}) \big|^2 \;
\sum_{{a_{n}}\in (a^{(n_0)}, \delta a)}  1  
= \big| \psi(a^{(n_0)}) \big|^2
(\delta a +1)
& {\rm for}  & 
\delta a \ll  \sigma_a \; ,  
\label{D a}   \\
\sum_{{\rm all} \; a_n}
|\langle a_n | \psi  \rangle_s|^2
=1 \; .
& {\rm for} &
\delta a \gg \sigma_a  
\label{D b}  \\
\end{array}
\right.
\label{D}
\end{eqnarray}
\label{N,D}
\end{subequations}

The conditional Wigner's formula, Eqs. (\ref{W(b|da)}),
(\ref{N(bm),D}),
then gives, in these two extreme cases
\begin{equation}
{\cal W}(b_m |a^{(n)}, \delta a)
\approx
\left\{
\begin{array}{ccccc}
\frac{\left|\sum_{a_n \in (a^{(n_0)}, \delta a)} 
\langle b_m |a_n  \rangle \right|^2 }{\delta a +1} \;,
&   
&{\rm for} 
& \delta a \ll  \sigma_a,  
& \hspace{1cm} {\rm (a)} \\
\left| \langle b_m | \psi \rangle_s  \right|^2 
\equiv |\psi_s (b_m)|^2 \; .
& &  
{\rm for} 
&\delta a \gg  \sigma_a,
& \hspace{1cm} {\rm (b)}
\end{array}
\right.
\label{cond_wigner_formula_an_bm}
\end{equation}
The resulting $b_m$ distribution of Eq. (\ref{cond_wigner_formula_an_bm}a)
is {\em independent of the system original state} $| \psi\rangle_s$;
{\em its width as a function of $\delta a$ is the UR} referred to in the previous section 
[see, e.g., Eq. (\ref{delta b}), or Eqs. (\ref{Delta b}) and
(\ref{varBpert=F(da)}) if the variance is well defined].
We shall be more concrete in the cases treated in Secs.
\ref{[A,B] arbitr}, \ref{periodic_model} and \ref{application to continuous position-momentum}
below.

As a check, 
i) when $\delta a \ll \sigma_a$
\begin{subequations}
\begin{eqnarray}
\sum_{b_m} {\cal W}(b_m | a^{(n_0)}, \delta a)
&=& \frac{1}{\delta a + 1} \;
\sum_{a_{n},a_{n'} \in (a^{(n_0)}, \delta a)}  
\langle a_{n'}|a_{n} \rangle \\
&=& \frac{1}{\delta a + 1} \;
\sum_{a_{n},a_{n'} \in (a^{(n_0)}, \delta a)}  
\delta_{n n'}
= \frac{1}{\delta a + 1} \times
\sum_{a_{n} \in (a^{(n_0)}, \delta a)}  1
= \frac{1}{\delta a + 1} (\delta a +1) =1 ;
\nonumber \\
\end{eqnarray}
\end{subequations}
ii) when $\delta a \gg \sigma_a$
\begin{eqnarray}
\sum_{b_m} {\cal W}(b_m | a^{(n_0)}, \delta a)
= \sum_{b_m}|\psi_s (b_m)|^2 
=1.
\end{eqnarray}


\subsection{The commutative case: $[\hat{A}, \hat{B}]=0$}.
\label{[A,B]=0}

As a particular situation, we investigate the consequence of
our two observables $\hat{A}$, $\hat{B}$ being {\em commutative}, i.e., $[B,A]=0$.
We assume $\delta a\ll \sigma_a$ and consider the two following cases:
 
1) Suppose $\hat{B}=\hat{A}$, i.e., after measuring $\hat{A}$, we measure the same observable again.
Then $|b_m \rangle = |a_m \rangle$ and $b_m=a_m$, and
Eq. (\ref{cond_wigner_formula_an_bm}a) gives
\begin{subequations}
\begin{eqnarray}
{\cal W}(b_m  |a_n \in (a^{(n_0)}, \delta a))
&\approx&
\frac{1}{\delta a + 1} \;
\Big| 
\sum_{{a_{n}}\in (a^{(n_0)}, \delta a)  }  
\langle b_m|a_n \rangle \Big|^2  
\label{W(b|da)da<<sigm_a A=B a}   \\
& \stackrel{|b_m\rangle = |a_m\rangle}{=} & 
\frac{1}{\delta a  + 1} \;
\Big| 
\sum_{{a_{n}}\in (a^{(n_0)}, \delta a)}  
\langle a_m|a_n \rangle \Big|^2   
\label{W(b|da)da<<sigm_a A=B b}  \\
&=& \frac{1}{\delta a + 1} \;
\Big| 
\sum_{{a_{n}}\in (a^{(n_0)}, \delta a)}  
\delta_{a_m, a_n}\Big|^2  
\label{W(b|da)da<<sigm_a A=B c}  \\
&=& \frac{1}{\delta a + 1} \;
\theta(a_{m}\in (a^{(n_0)}, \delta a))
\label{W(b|da)da<<sigm_a A=B d}  \\
& \stackrel{b_m = a_m}{=} &
\frac{1}{\delta a + 1} \;
\theta(b_{m}\in (a^{(n_0)}, \delta a))\; ,
\label{W(b|da)da<<sigm_a A=B e}
\end{eqnarray}
\label{W(b|da)da<<sigm_a A=B}
\end{subequations}
where we have defined the function 
$\theta(a_{m}\in (a^{(n_0)}, \delta a)) \equiv 1$ if 
$a_{m}\in (a^{(n_0)}, \delta a)$, 
and $\theta(a_{m}\in (a^{(n_0)}, \delta a)) \equiv 0$ if
$a_{m} \not\in (a^{(n_0)}, \delta a)$.
This is an understandable result, giving 
\begin{equation}
\delta b = \delta a \; ,
\label{db=da for A=B}
\end{equation}
$\delta b$ being the number of states
$b_m \neq b_{m_0}$ that the measurement cannot distinguish from
$b_{m_0}$.

2) Suppose
\begin{subequations}
\begin{eqnarray}
\hat{B}\neq \hat{A}, \\
{\rm but} \;\;  [\hat{B}, \hat{A}] = 0 .
\end{eqnarray}
\label{BneqA [A,B]=0}
\end{subequations}

By a well-known textbook argument, we have the following results:

i) If every $a_n$ is {\em not degenerate} $\Rightarrow$ every eigenfunction of $\hat{A}$ is an eigenfunction of $\hat{B}$.

ii) If every $b_m$ is {\em not degenerate} $\Rightarrow$ every eigenfunction of $\hat{B}$ is an eigenfunction of $\hat{A}$.

\begin{figure}[!ht]
\centering
\includegraphics[width=6cm,height=5cm]{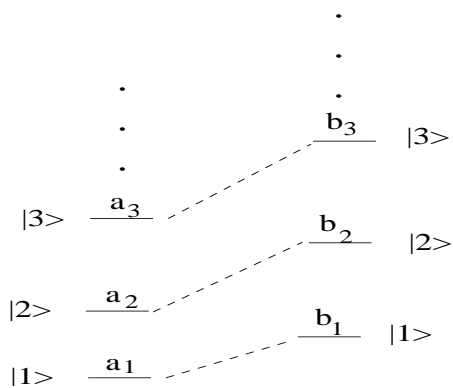}
\caption{
The non-degenerate spectra of two, in general different, commuting operators, $[\hat{A},\hat{B}]=0$.
All the $a_n$s are different from one another;
also, all the $b_m$s are different from one another.
}
\label{b=f(a)}
\end{figure}

\noindent
Thus, since {\em both spectra are assumed to be non-degenerate}, 
given $a_n$ we have, uniquely, one $b_n$, and vice-versa.
This is illustrated schematically in Fig. \ref{b=f(a)}.
This association $a_n \Leftrightarrow b_n$, $\forall n$,
can be described as a function $b_n = f(a_n)$,
and  hence $\hat{B}=f(\hat{A})$.
As an example, $\hat{A}$ could be a 1D harmonic oscillator Hamiltonian,
${\hat A}= H^{1D}_{HO}(\omega)$, and 
${\hat B}= [H^{1D}_{HO}(\omega)]^2$,

We again assume $\delta a \ll \sigma_a$. The consequences for Wigner's rule (\ref{cond_wigner_formula_an_bm}a)
are as follows.
Since we have 
$b_n = f(a_n)$
and $|b_n\rangle = |a_n \rangle$  (see Fig. \ref{b=f(a)}),
Eq. (\ref{cond_wigner_formula_an_bm}a) gives
\begin{subequations}
\begin{eqnarray}
{\cal W}(b_m  |a_n \in (a^{(n_0)}, \delta a))
&\approx&
\frac{1}{\delta a + 1} \;
\Big| 
\sum_{{a_{n}}\in (a^{(n_0)}, \delta a)  }  
\langle b_m|a_n \rangle \Big|^2  
\label{W(b|da)da<<sigm_a [A,B]=0 a}    \\
& \stackrel{|b_m\rangle = |a_m\rangle}{=} &
 \frac{1}{\delta a  + 1} \;
\Big| 
\sum_{{a_{n}}\in (a^{(n_0)}, \delta a)}  
\langle a_m|a_n \rangle \Big|^2   
\label{W(b|da)da<<sigm_a [A,B]=0 b}  \\
&=& \frac{1}{\delta a + 1} \;
\Big| 
\sum_{{a_{n}}\in (a^{(n_0)}, \delta a)}  
\delta_{a_m, a_n}\Big|^2  
\label{W(b|da)da<<sigm_a [A,B]=0 c}  \\
&=& \frac{1}{\delta a + 1} \;
\theta(a_{m}\in (a^{(n_0)}, \delta a))
\label{W(b|da)da<<sigm_a [A,B]=0 d}  \\
&=& \frac{1}{\delta a + 1} \;
\theta(f(a_{m})\in f(a^{(n_0)}, \delta a))
\label{W(b|da)da<<sigm_a [A,B]=0 e}  \\
& \stackrel{f(a_m) = b_m}{=} &
\frac{1}{\delta a + 1} \;
\theta(b_{m}\in f(a^{(n_0)}, \delta a))
\label{W(b|da)da<<sigm_a [A,B]=0 f}                 
\end{eqnarray}
\label{W(b|da)da<<sigm_a [A,B]=0}
\end{subequations}
From the definition (\ref{delta b}), this gives the result 
\begin{equation}
\delta b = \delta a \; ,
\label{db=da for a neq B [A,B]=0}
\end{equation}
just as in case 1) above.

We may also describe the present case (\ref{BneqA [A,B]=0})
in terms of the first and second moment and variance of the $b_m$ distribution, as in Eq. (\ref{Delta b}).
First, the matrix elements of $\hat{A}$ and $\hat{B}$ are
\begin{subequations}
\begin{eqnarray}
\langle a_{n'}|\hat{A}|a_{n}\rangle 
&=& a_n \delta_{n n'} \; ,  
\label{m.els. of A}  \\
\langle a_{n'}|\hat{B}|a_{n}\rangle 
&=& b_n \delta_{n n'}
= f(a_n) \delta_{n n'} \; .
\label{m.els. of B}
\end{eqnarray}
\label{m.els. of A and B}
\end{subequations}
Multiplying Eq. (\ref{W(b|da)da<<sigm_a [A,B]=0 a}) by $b_m$ and $b_m^2$ and summing over $m$,
and using the matrix elements (\ref{m.els. of B}), 
we find the first and second moments of the $b_m$ distribution as
\begin{subequations}
\begin{eqnarray}
E(B|a_n \in (a^{(n_0)}, \delta a))
&=&\frac{1}{\delta a + 1} \;
\sum_{a_{n},a_{n'} \in (a^{(n_0)}, \delta a)}  
\langle a_{n'}|\hat{B}|a_{n} \rangle
= \frac{1}{\delta a + 1} \;
\sum_{a_{n} \in (a^{(n_0)}, \delta a)}  
f(a_n)\; ,
\label{E(B|da)} \nonumber \\
 \\
E(B^2|a_n \in (a^{(n_0)}, \delta a))
&=&\frac{1}{\delta a + 1} \;
\sum_{a_{n},a_{n'} \in (a^{(n_0)}, \delta a)}  
\langle a_{n'}|\hat{B}^2|a_{n} \rangle
= \frac{1}{\delta a + 1} \;
\sum_{a_{n} \in (a^{(n_0)}, \delta a)}  
f^2(a_n) \; .
\label{E2(B|da)} \nonumber \\
\end{eqnarray}
The variance
\begin{eqnarray}
(\Delta \hat{B})^2
\equiv {\rm var}(B|a_n \in (a^{(n_0)})
&=& 
\left[
\frac{1}{\delta a + 1} \;
\sum_{a_{n} \in (a^{(n_0)}, \delta a)}  
f^2(a_n) \right]
- \left[
\frac{1}{\delta a + 1} \;
\sum_{a_{n} \in (a^{(n_0)}, \delta a)}  
f(a_n)
\right]^2 \; ,
\label{var(B|da)}
\nonumber \\ 
\end{eqnarray}
\end{subequations}
is a measure, alternative to $(\delta b)^2$, of the width of the $b_m$ distribution; it depends on the actual values of the spectral quantities $b_m=f(a_m)$.

\subsection{The general case: $[\hat{A}, \hat{B}]$ arbitrary}.
\label{[A,B] arbitr}

\subsubsection{A model for the unitary matrices 
$|| \langle b_m | a_n \rangle ||$ of Eq. (\ref{cond_wigner_formula_an_bm}a)}
\label{model for unitary matrices}

We assume 
i) the $\psi_s(a_n)$ components of the system wavefunction to be centered at the same value $a^{(n_0)}$ around which the first low-resolution measurement is performed; 
ii) $\delta a \ll \sigma_a$, so that Eq. (\ref{cond_wigner_formula_an_bm}a)
applies
(see Fig. \ref{fig1}).

\paragraph{Preliminaries}
\label{prelim}

We assume all along that the spectra $a_n$ and $b_m$ are {\em non-degenerate}.

i) Eqs. (\ref{db=da for A=B}) and (\ref{db=da for a neq B [A,B]=0})
indicate that
\begin{subequations}
\begin{eqnarray}
&& [A,B]=0    \\
{\rm implies}  
&& \delta b = \delta a \; .
\end{eqnarray}
\label{Da=Db}
\end{subequations}

         ii) For Schwinger's discrete periodic model 
\cite{schwinger} of 
Sec. \ref{periodic_model} ahead,
Eq. (\ref{Deltap.Deltaq ge N}) gives
\begin{subequations}
\begin{eqnarray}
{\rm For}: &&  \hat{A}=\hat{p}, \;\;\;  \hat{B}={q}     \\
&& \delta q  \sim N/\delta p
\label{Dp=1/Dq}
\end{eqnarray}
\end{subequations}
and for the continuous position-momentum case of 
Sec. \ref{application to continuous position-momentum}
ahead, Eq. (\ref{cond_wigner_formula_posi_mom 3}) establishes that
\begin{subequations}
\begin{eqnarray}
{\rm For}: \;\;\;   \hat{A}=\hat{p}, \;\;\;  \hat{B}=\hat{q}     \\
 \delta q  \sim 1/\delta p
\label{Dp=1/Dq cont}
\end{eqnarray}    
\label{continuous pos and mom}  
\end{subequations}

The results i) and ii) are very different:
i)  in (\ref{Da=Db}), the uncertainty on the RHS is in the numerator;  
ii)  in (\ref{Dp=1/Dq}) and (\ref{Dp=1/Dq cont}),  
the uncertainty on the RHS  is in the  denominator.
Of course, these are two extreme cases.
If $U$ denotes the unitary matrix 
$\left|| \langle b_m|a_n \rangle \right||$ appearing in Eq. (3.5a),
when we go from case i) to case ii) above, we go from a diagonal $U$ to a ``full" $U$.
For an intermediate case, $U$  is ``intermediate", depending on
$[\hat{A}, \hat{B}]$.
We need to write $U$ with ``intermediate properties".
In order to parametrize such a $U$,
we propose a model for the {\em continuous case}, based on
the unitary matrix of Eq. (32) in Ref. \cite{khanna_et_al_2012}.             

\paragraph{The model}

Eq. (25) of Ref. \cite{khanna_et_al_2012} contemplates the canonical transformation
leading from the canonical variables $\hat{x}$, $\hat{p}$ to the new canonical variables
\begin{subequations}
\begin{eqnarray}
{\hat X}_{\theta}  &=& \cos \theta  \; {\hat x} + \sin \theta \; {\hat p}  \; , \\
{\hat P}_{\theta} &=& - \sin \theta  \; {\hat x} + \cos \theta \; {\hat p} \; .
\end{eqnarray}
\label{X, P (theta)}
\end{subequations}
The operators ${\hat X}_{\theta}$ and ${\hat P}_{\theta}$
are canonically conjugate, i.e., 
$[{\hat X}_{\theta}, {\hat P}_{\theta}] = i$,
just as the original operators ${\hat x}$ and ${\hat p}$.

We identify the operator $\hat{A}$ of the previous sections with the present operator ${\hat X}_{\theta}$, and the operator $\hat{B}$
with the present operator ${\hat x}$.
This identification and that of the corresponding eigenstates is then
\begin{subequations}
\begin{eqnarray}
&& \hat{A}_{\theta} \equiv {\hat X}_{\theta} \;\;\;
\Longrightarrow \;\;\; 
|a \rangle = | x', \theta \rangle  
= U^{\dagger}(\theta)| x' \rangle 
\;\; \Rightarrow \;\; \hat{A}_{\theta} | x', \theta \rangle   = x' | x', \theta \rangle \; , \\
&& \hat{B} = \hat{x} \;\;\;\;\;\;\;
\Longrightarrow \;\; 
|b \rangle  =  | x \rangle
\;\; \Rightarrow \;\; \hat{B} | x \rangle   = x | x \rangle \; .
\end{eqnarray}
\label{A=X, B=x}
\end{subequations}
The commutator of the two observables $\hat{B}$  and 
$\hat{A}_{\theta}$ is given by
\begin{eqnarray}
[\hat{B}, \hat{A}_{\theta}] 
= [\hat{x},\; \cos \theta  \; {\hat x} + \sin \theta \; {\hat p} ]
= i \sin \theta ,
\end{eqnarray}
so the parameter $\theta$ can be viewed as a measure of the commutator.

Just as above, we shall study the conditioned Wigner formula  (\ref{W(b|da)})  for the probability of finding $x$, conditioned on having found $x'$ in the interval $\delta x'$ around the value $x'=0$.
We consider  the case $\delta x' \ll \sigma_{x'}$   illustrated in Fig. \ref{fig1}, 
in which the initial system wavefunction $\psi_s(x')$ has a large spread $\sigma_{x'}$
compared with the resolution $\delta x'$ of the first measurement,
and is centered at the same value, $x'=0$, around which the $x'$s are looked for in the first measurement.

The overlap $\langle b_m | a_n \rangle$ appearing in Eq. 
(\ref{cond_wigner_formula_an_bm}a)
above now corresponds to
[from Eq. (32) of Ref. \cite{khanna_et_al_2012}]
\begin{subequations}
\begin{eqnarray}
\hspace{1cm} \langle b_m | a_n \rangle 
&\Rightarrow& \langle  x | x', \theta  \rangle 
\label{<b|a>1}     \\
&=& \langle x  | U^{\dagger}(\theta) |   x'  \rangle 
 = \frac{  {\rm e}   ^{  i(\frac{\pi}{4} - \frac{\theta}{2})} }   
 {\sqrt{2 \pi |\sin(\theta)|}}
 {\rm e}^{-\frac{i}{2 \sin \theta} 
 [(x^2 + x'^{2})\cos \theta - 2 x x']}\;,
\label{<b|a>2}
\end{eqnarray}
\label{<b|a>}
\end{subequations}
($0 \le \theta \le \pi$).
Then the sum $S(b_m)$ appearing in Eq. (3.5a) will be denoted by
$S(x; x' \in (0,\delta x'))$, i.e.,
\begin{subequations}
\begin{eqnarray}
 S(b_m) &=& \sum_{a_n  \in ( a^{(n_0)}, \delta a)} \langle b_m | a_n \rangle 
\label{S(b) a}  \\
 \Rightarrow S(x; x' \in (0,\delta x')) &=& \int_{-\delta x'/2}^{\delta x'/2} \langle  x | x', \theta  \rangle dx' 
= \int_{-\delta x'/2}^{\delta x'/2} \langle x  | U^{\dagger}(\theta) |   x'  \rangle dx'         
\label{S(x) b} \\
&=& f(\theta)
\int_{-\delta x'/2}^{\delta x'/2}
{\rm e}^{-\frac{i}{2 \sin \theta} [(x^2 + x'^{2})\cos \theta - 2 x x']} dx'  
\label{S(x) c}   \\
&=& f(\theta)\frac{\sqrt{\pi}}{2} (-1)^{3/4}{\rm e}^{\frac{i}{2}x^2 \tan(\theta)}
\left[
{\rm erf}\frac{(-1)^{1/4}(x-\frac12 \delta x' \cos(\theta))}{\sqrt{\sin(2 \theta)}}
\right.  
\nonumber \\
&& \hspace{1cm} \left.
- {\rm erf}\frac{(-1)^{1/4}(x+\frac12 \delta x' \cos(\theta))}{\sqrt{\sin(2 \theta)}}
\right]
\sec(\theta)\sqrt{\sin(2 \theta)}
\label{S(x) d}
\end{eqnarray}
\label{S(x)}
\end{subequations}
$f(\theta)$ being the prefactor in Eq. (\ref{<b|a>2});
${\rm erf}(z)$ denotes the error function.
The powers of $(-1)$ are understood to represent principal values:
thus $(-1)^{1/4}= {\rm e}^{i\pi/4} = (1+i)/\sqrt{2}$ and
$(-1)^{3/4}= {\rm e}^{i 3 \pi/4} =(-1+i)/\sqrt{2}$.
The result (\ref{S(x) d}) was obtained using Mathematica.

For $\delta x' \ll \sigma_{x'}$, the conditional Wigner formula of Eq. (\ref{cond_wigner_formula_an_bm}a)
gives
\begin{eqnarray}
{\cal W}(x|x'\in \delta x'; \theta)
&=& \frac{1}{\delta x'}
\left|\int_{-\delta x'/2}^{\delta x' /2}
\langle  x | x', \theta  \rangle dx'\right|^2       \\
&=& \frac{\left| S(x; x' \in (0,\delta x')) \right|^2}{\delta x'}
\label{wigner formula x'(theta) x}
\end{eqnarray}
As a check, we verify the normalization of the conditioned Wigner formula of 
Eq. (\ref{wigner formula x'(theta) x}):
\begin{eqnarray}
\int_{-\infty}^{\infty}{\cal W}(x|x'\in \delta x'; \theta) dx
&=& \int_{-\infty}^{\infty} \frac{dx}{\delta x'}
\int_{-\delta x'/2}^{\delta x' /2} dx'
\int_{-\delta x'/2}^{\delta x' /2} dx''
\langle x'', \theta  |  x \rangle
\langle x | x', \theta \rangle 
\nonumber\\
&=&\frac{1}{\delta x'}
\int_{-\delta x'/2}^{\delta x' /2} dx'
\int_{-\delta x'/2}^{\delta x' /2} dx''
\; \delta(x''-x') 
\nonumber \\
&=&\frac{1}{\delta x'}
\int_{-\delta x'/2}^{\delta x' /2} dx'
\; \theta(x'  \in \delta x')
\nonumber \\
&=&\frac{1}{\delta x'} \delta x' =1,
\label{norm wigner formula x'(theta) x}
\end{eqnarray}
as expected.

Since the quantity 
$S(x; x' \in (0,\delta x'))$ found in Eq. (\ref{S(x) d})
is not easy to be handled analytically, it
was plotted as function of $x$ for various values of $\delta x'$ by means of Mathematica: the following numerical results were obtained:
\begin{subequations}
\begin{eqnarray}
\delta x' &=& 1 \\
\theta &=&  0.001; \hspace{5mm} [\hat{x}, \hat{x'}]= 0.001 i \; ; 
\hspace{5mm}  \delta x \approx 1 \approx \delta x'         \\
\theta &=& \pi/4; \hspace{7mm} [\hat{x}, \hat{x'}]= \frac{i}{\sqrt{2}}\; \;
\hspace{11mm}  \delta x \approx 8.4            \\
\theta &=& \pi/2, \hspace{7mm} [\hat{x}, \hat{x'}]= i \; ;
\hspace{15mm}  \delta x \approx 12 \approx 4 \pi / \delta x'
\end{eqnarray}
\end{subequations}
and
\begin{subequations}
\begin{eqnarray}
\delta x' &=& 2 \\
\theta &=&  0.01; \hspace{5mm} [\hat{x}, \hat{x'}]= 0.01 i \; ; 
\hspace{5mm}  \delta x \approx 2 \approx \delta x'         \\
\theta &=& \pi/4; \hspace{7mm} [\hat{x}, \hat{x'}]= \frac{i}{\sqrt{2}}\; \;
\hspace{11mm}  \delta x \approx 4            \\
\theta &=& \pi/2, \hspace{7mm} [\hat{x}, \hat{x'}]= i \; ;
\hspace{15mm}  \delta x \approx 6 \approx 4 \pi / \delta x'
\end{eqnarray}
\end{subequations}
These numerical results are consistent with our earlier remarks in 
Eqs. (\ref{Da=Db}) and (\ref{continuous pos and mom}).

Since the erf is not easy to handle analytically, 
we examine a model for it, which can be integrated in an elementary way, as follows.
We replace Eq. (\ref{S(x) c}) by the following one:
\begin{eqnarray}
S^{model}(x; x' \in (0,\delta x'))
\equiv f(\theta)
\int_{-\infty}^{\infty}
{\rm e}^{- \frac{ {x'}^{2} }{ c\left(\frac{\delta x'}{2}\right)^2 } }
{\rm e}^{-\frac{i}{2 \sin \theta} [(x^2 + x'^{2})\cos \theta - 2 x x']} dx',
\label{S(x)model 1}
\end{eqnarray}
in which the sharp limits of integration  
$-\delta x'/2, \delta x'/2$ in (\ref{S(x) c}) were replaced
by $-\infty, +\infty$, and the factor 
${\rm e}^{- \frac{ {x'}^{2} }{ c\left(\frac{\delta x'}{2}\right)^2 } }$ 
was added to the integrand, so as to give more weight to the interval
$-\delta x'/2, \delta x'/2$.
The factor $c$ may be a useful adjustable parameter.
The result of the integration in Eq. (\ref{S(x)model 1}) is 
\begin{eqnarray}
|S^{model}(x; x' \in (0,\delta x'))|^2
= \frac{c}{|\sin \theta|}
\frac{\frac12(\frac{\delta x'}{2})^2}
{\sqrt{1 + (\frac{c}{2})^2(\frac{\delta x'}{2})^4 \cot^2 \theta}}
{\rm e}^{ 
- \frac{x^2}{\sin^2 \theta}  
\frac{\frac{c}{2}(\frac{\delta x'}{2})^2 }
{1 + (\frac{c}{2})^2(\frac{\delta x'}{2})^4 \cot^2 \theta} 
}
\label{S(x)model 2}
\end{eqnarray}
If we define $\delta x /2$ as the value of $x$ for which 
$|S^{model}(x; {\delta x'})|^2$ decays to $1/e$, we have
\begin{eqnarray}
\frac{c}{2}\frac{(\frac{\delta x}{2})^2}{\sin^2 \theta}  \;
\frac{(\frac{\delta x'}{2})^2 }
{1 + (\frac{c}{2})^2(\frac{\delta x'}{2})^4 \cot^2 \theta} = 1 ,
\label{Dx 1}
\end{eqnarray}
or 
\begin{eqnarray}
\frac{c}{2}(\delta x')^2 (\delta x)^2
= 16 \left|[\hat{X}_{\theta},\hat{x}]\right|^2 + \left(\frac{c}{2}\right)^2 
\left(1 - \left|[\hat{X}_{\theta},\hat{x}]\right|^2\right)(\delta x')^4
\label{DxDx'}
\end{eqnarray}
It may be convenient to choose $c=2$, so that Eq. (\ref{DxDx'}) gives
\begin{eqnarray}
(\delta x')^2 (\delta x)^2
= 16 \left|[\hat{X}_{\theta},\hat{x}]\right|^2 + 
\left(1 - \left|[\hat{X}_{\theta},\hat{x}]\right|^2\right)(\delta x')^4 \; .
\label{DxDx' 2}
\end{eqnarray}
Particular cases are:
\begin{subequations}
\begin{eqnarray}
&&\theta \to 0, \;\;\; \hat{X}_{\theta} \to \hat{x}, \;\;\; \hat{B}= \hat{x}, 
\;\;\; \left|[\hat{X}_{\theta}, \hat{x}]\right| \to  0
\nonumber \\
&& \hspace{1cm} (\delta x)^2 = (\delta x')^2,
\; {\rm as \; in \; Eq.} (\ref{Da=Db}) \\
&&\theta = \pi/2, \;\;\;  \hat{X}_{\theta}=\hat{p}, \;\;\; \hat{B}= \hat{x}, 
\;\;\; \left|[\hat{X}_{\theta}, \hat{x}]\right|  = 1,  
\nonumber \\
&& \hspace{1cm}(\delta x')^2 (\delta x)^2 = 16 ,
\; {\rm as \; in \; Eq.} \; (\ref{Dp=1/Dq cont}).     
\label{DxDx' 2 part cases}
\end{eqnarray}
\end{subequations}

Consequences of Eq. (\ref{DxDx' 2}) are the inequalities
\begin{subequations}
\begin{eqnarray}
&& (\delta x')^2 (\delta x)^2
\ge 16 |[\hat{X}_{\theta}, \hat{x}]|^2 = 16 \sin^2 \theta \; 
\Rightarrow
\left\{
\begin{array}{cccc}
(\delta x')^2 (\delta x)^2 \ge 0,
& {\rm if} & 
\theta \to 0 \;\;\;\;\;(a1)  
\label{} \\
(\delta x')^2 (\delta x)^2 \ge 16, 
& {\rm if} &
\theta = \pi/2. \;\;(a2)
\label{}
\end{array}
\right.
\label{inequality a}   \\
&& \;\;\;\;\;  {\rm If} \; \delta x' \neq 0 \;\;\; \Rightarrow 
\nonumber \\
&& (\delta x)^2 
\ge \left(1 - |[\hat{X}_{\theta}, \hat{x}]|^2\right)(\delta x')^2
= (\cos^2 \theta)  (\delta x')^2  
\nonumber \\
&& \hspace{5cm} \Rightarrow
\left\{
\begin{array}{cccc}
(\delta x)^2 \ge  (\delta x')^2 ,
& {\rm if} & 
\theta \to 0 , \;\;\;\;\;\;\;\;\;(b1) \\
(\delta x)^2 \ge 0, 
& {\rm if} &
\theta = \pi/2,  \;\;\;\;\;\;(b2)
\end{array} .
\right.
\label{inequality b}
\end{eqnarray}
\label{inequalities}
\end{subequations}
Inequality 
(\ref{inequality a}) (before the curly bracket) is similar to the standard Robertson inequality
[\onlinecite{robertson}], except that here there is {\em no dependence left of the original system 
wavefunction, due to the assumption $\delta x' \ll \sigma_{x'}$}.
(The particular case (\ref{inequality a}1) is an obvious result.)
Inequality (\ref{inequality b}1) is consistent with Eqs. (\ref{Da=Db}).
(Inequality (\ref{inequality b}2) is an obvious result.)

\section{Periodic model for the matrix element $\langle b_m |a_n\rangle$}
\label{periodic_model}

The purpose of this section is to illustrate the UR
in a specific example, in which we can 
compute explicitly the above expressions for the conditional Wigner formula,
Eq. (\ref{cond_wigner_formula_an_bm}a);
we shall consider a particular model for the matrix element 
$\langle b_m |a_n\rangle$ that we now describe.
We assume, as always in this paper, the situation of strong coupling between the system and the first probe. 
We only consider the case $\delta a \ll \sigma_a$;
we assume that
the system wavefunction $\psi_s(a_n)$ is centered at the same value $a^{(n_0)}$ around which the first low-resolution measurement is performed, and that it has a large spread compared with the resolution $\delta a$ of the first measurement 
[Eq. (\ref{cond_wigner_formula_an_bm}a) applies;
see Fig. \ref{fig1}].

In an $N$-dimensional Hilbert space we consider Schwinger's operators
\cite{schwinger} with a periodic structure, as summarized in  App. \ref{schwinger}.
We define the usual {\em position-like} and {\em momentum-like} operators 
and identify the $| a_n\rangle$
and $| b_m\rangle$ eigenstates and the corresponding projectors of Eqs. (\ref{sp repr A,B}), as
\begin{subequations}
\begin{eqnarray}
&& |a_n\rangle
=|p\rangle, \;\;\;\;\;\;   \mathbb{P}_p = |p\rangle \langle p |, 
\;\;\;\;\;\;  p=0,1, \cdots, N-1  ,          
\label{p schw}   \\
&& |b_m\rangle
=|q\rangle , \;\;\;\;\;\;   \mathbb{P}_q = |q\rangle \langle q |, 
\;\;\;\;\;\; q=0,1, \cdots, N-1  ,          
\end{eqnarray}
the two bases being related as
\begin{eqnarray}
\hspace{2cm} |p\rangle = \sum_{q=0}^{N-1}
\frac{   e^{  \frac{2\pi i } {N}  p q    } }   { \sqrt{N} } 
 |q\rangle \; .
\label{p schw}
\end{eqnarray}
\label{p q schwinger}
\end{subequations}

Eq. (\ref{sp repr A delta a 0}) now corresponds to the {\em low-resolution momentum-like operator and momentum-like projector}, i.e.,
\begin{subequations}
\begin{eqnarray}
\hat{p}^{\delta p}
&=& \sum_{n=1}^{n_{max}} p^{(n)} \; \mathbb{P}_{p^{(n)} }^{\delta p},
\;\;\;\; n = 1, \cdots, n_{max}= \frac{N}{\delta p + 1}
\label{pdp per. model} \\
\mathbb{P}_{p^{(n)}}^{\delta p}
&=&\sum_{p'=p^{(n)} - \frac{\delta p}{2}}^{p^{(n)} + \frac{\delta p }{2}} 
\mathbb{P}_{p'} 
\;\;\;\;\; (\delta p +1 \; {\rm terms}) \; .
\label{Pdp per. model}
\end{eqnarray}
\label{per. model}
\end{subequations}

For clarity in the interpretation of 
$\delta p$, $n_{max}$, $N$, and related quantities,
we repeat the following points, following the assumptions around 
Eqs. (\ref{sp repr A,B 0}):

i) the various momentum intervals are {\em disjoint}, 
have a width $\delta p$, and contain $\delta p +1$ levels each;

ii) we assume 
\begin{subequations}
\begin{eqnarray}
\delta p &=& {\rm even} ;
\label{dp even} 
\end{eqnarray}

iii) $N$ is such that 
\begin{eqnarray}
N&=& n_{max}(\delta p +1),
\label{N,nmax,dp}
\end{eqnarray}
\label{dp, N}
\end{subequations}
so that it contains an integral number $n_{max}$ of disjoint $\delta p$ intervals,
centered at $p^{(n)}=p^{(1)},  p^{(2)}, \cdots, p^{(n_{max})}$, i.e.,
\begin{subequations}
\begin{eqnarray}
p^{(1)} &=& \frac{\delta p}{2} \; , \\
p^{(2)}  &=& \frac{\delta p}{2} + (\delta p + 1) \; ,\\
\cdots  \\
p^{(n_{max})}
&=& \frac{\delta p}{2} + (n_{max}-1)(\delta p + 1)
=N-1 - \frac{\delta p}{2} \; .
\end{eqnarray}
\end{subequations}

iii) As a result, $N$ {\em cannot be a prime number}.
For the present analysis, 
in which we only contemplate the two bases $p$ and $q$, 
a prime-number requirement for $N$ is not needed.
(For instance, the need for $1/2 \;{\rm mod} N$ never arises.)

Eq. (\ref{sp repr B 0}) corresponds now to the
full-resolution position operator, for which we take
\begin{eqnarray}
\mathbb{P}_q = |q\rangle \langle q | 
\;\;\;\;\;\; \Rightarrow \;\;\; 
\hat{q} = \sum_{q=0}^{N-1}
q \; \mathbb{P}_q \; .
\end{eqnarray}

\begin{figure}[!ht]
\centering
\includegraphics[width=12cm,height=6cm]{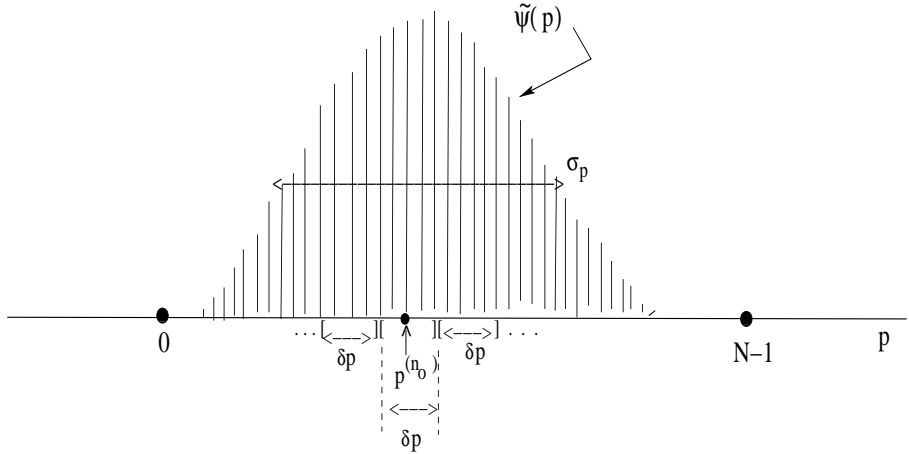}
\caption{
Schematic representation of the momentum $p$ components 
$\langle p | \psi \rangle_s = \widetilde{\psi}_s(p)$ of the system original state, for the case $\delta p \ll \sigma_p \lesssim N$,
and $N \gg 1$.
The system wavefunction components $\widetilde{\psi}_s(p)$ are assumed centered at the same value $p^{(n_0)}$ around which the first low-resolution $p$-measurement is performed.
}
\label{tilde_psi(p)}
\end{figure}

The sum appearing in
Eq. (\ref{cond_wigner_formula_an_bm}a), to be called
$S_{p^{(n_0)}, \delta p}(q)$ in the present case, 
is relevant for the case 
$\delta p \ll \sigma_p \lesssim N$, 
which implies $\delta p \ll N$, and is meaningful when $N \gg 1$:
see Fig. \ref{tilde_psi(p)}.
The sum $S_{p^{(n_0)}, \delta p}(q)$ contains
$\delta p + 1$ terms and can be computed explicitly
in the present model, giving
\begin{subequations}
\begin{eqnarray}
S_{p^{(n_0)}, \delta p}(q) 
&\equiv& \sum_{p=p^{(n_0)}-\frac{\delta p}{2}}
^{p^{(n_0)} + \frac{\delta p}{2}} 
\frac{e^{  \frac{2\pi i } {N}  p q    } }   {\sqrt{N}} \; ,    \\
&=&\frac{\omega^{ q  p^{(n_0)}}}{\sqrt{N}}\frac{\sin \frac{\pi q (\delta p + 1)}{N} }{\sin \frac{\pi q}{N}} ,
\;\;\;\;\;\;\;\; \omega \equiv {\rm e}^{\frac{2 \pi i}{N}} \; .
\end{eqnarray}
\end{subequations}
As an illustration, let $\delta p =0$.
Then
\begin{subequations}
\begin{eqnarray}
&&  n_{max} = \frac{N}{\delta p + 1} = N ,
\end{eqnarray}
and the $p^{(n)}$ are given by
\begin{eqnarray}
&&  p^{(1)}=0, \; p^{(2)}=1, \cdots, p^{(n_{max})}=N-1  .
\end{eqnarray}
We have $N$ disjoint intervals of width 0, i.e., containing just one level each.
Thus
\begin{eqnarray}
&& S_{p^{(n_0)}, \delta p}(q) 
= \frac{e^{  \frac{2\pi i } {N}  p^{(n_0)} q}  }   { \sqrt{N} } .
\end{eqnarray}
\end{subequations}

Going back to $\delta p$ arbitrary, Eq. (\ref{cond_wigner_formula_an_bm}a), we find
\begin{eqnarray}
&& {\cal W}(q |p \in (p^{(n)}, \delta p))
\approx
\frac{1}{N (\delta p + 1)} \;
\left(
\frac{ \sin \frac{\pi q (\delta p + 1)}{N}  }
 { \sin \frac{\pi q }{N} }
\right)^2  \; ,  \;\;\;\;\; q=0, \cdots, N-1 \; .
\label{W(q|)}   
\end{eqnarray}
The {\em width of this distribution} is the {\em UR} referred to above in Eq. (\ref{delta b}).

Again, for the particular case $\delta p = 0$, Eq. (\ref{W(q|)}) reduces to
\begin{eqnarray}
{\cal W}(q |p \in (p^{(n)}, \delta p = 0))
=\frac{1}{N}\left( \frac{\sin \frac{\pi q}{N}}{\sin \frac{\pi q}{N}} \right)^2 
= \frac1N \; ,  \;\;\;\;\; q=0, \cdots, N-1 \;,  
\label{W(q|) dp=0}   
\end{eqnarray}
and we define
\begin{eqnarray}
\delta q \equiv N-1 \; ,
\label{dq for dp=0}   
\end{eqnarray}
as the length of the $q$ span over which 
the distribution (\ref{W(q|) dp=0}) equals $1/N$.

\begin{figure}[t]
\centering
\includegraphics[width=12cm,height=6cm]{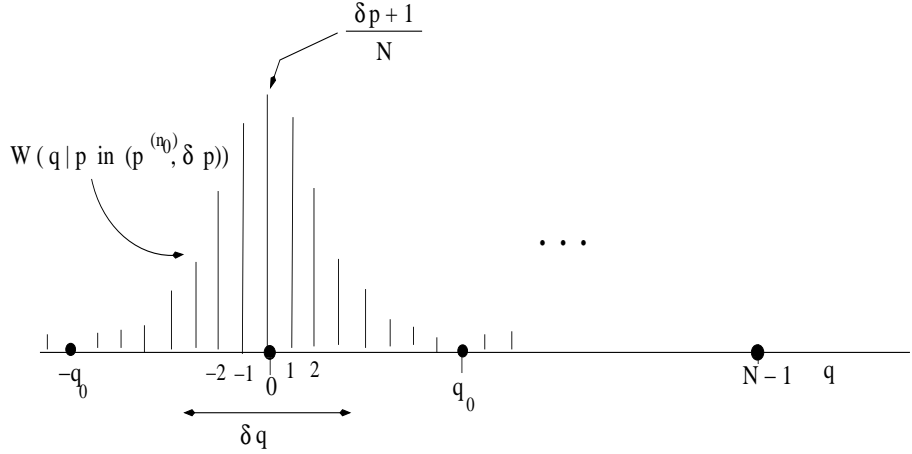}
\caption{
Schematic representation of the distribution
${\cal W}(q |p \in (p^{(n)}, \delta p))$ of Eq. (\ref{W(q|)}).
The first zero of this distribution, $q_0$, is indicated, as well as its width $\delta q$
(see text above Eq. (\ref{Deltap.Deltaq ge N})).
}
\label{W(q | p in dp)}
\end{figure}

Away from the extreme case $\delta p =0$, we fix our attention 
on the first zero of the distribution (\ref{W(q|)}), i.e., 
when $\frac{\pi q_0 (\delta p + 1)}{N} = \pi$, 
giving $q_0\equiv \frac{N}{\delta p + 1}$;
we define the width $\delta q$ of the $q$ distribution to be a function of $q_0$,
like $\delta q \equiv \alpha q_0$, that reduces to $\delta q = N-1$
when $\delta p = 0$.
We find $\alpha =(N-1)/N$, so that 
$\delta q = (N-1)/(\delta p +1)$, and thus
\begin{equation}
\delta q \cdot (\delta p + 1) =  N - 1 \; .
\label{Deltap.Deltaq ge N}
\end{equation}
The various quantities are illustrated in Fig. \ref{W(q | p in dp)}.

It is useful to compare Schwinger's model with the case of a free particle in 1D with periodic boundary conditions in the interval $[0, L]$ for the position 
$x$: in the latter case, $p_n = \frac{2 \pi}{L} \times (0, 1, \cdots)$, and
we get $\delta x \cdot \delta p \sim 1$.
In contrast, the equivalent of the factor $1/L$ appearing in $p_n$
does not appear 
in Schwinger's picture (where it would be $1/N$), where $p=0,1, \cdots$,
and we thus get the relation (\ref{Deltap.Deltaq ge N}), with 
the RHS as shown.

We end this section 
verifying the Robertson relation (\ref{robertson}) 
for the state $| \psi^{(p^{(n_0)}, \delta p)}  \rangle$, 
i.e.,
\begin{equation}
({\rm var} \hat{B})_{| \psi^{(p^{(n_0)}, \delta p)}  \rangle}
\ge \frac14 \frac
{ \left|
\left< \left[  \hat{A}, \hat{B}  \right]  \right>_{| \psi^{(p^{(n_0)}, \delta p)}  \rangle }\right|^2 }
{\left( {\rm var} \hat{A}  \right)_{| \psi^{(p^{(n_0)}, \delta p)}  \rangle }  } \; ,
\label{robertson 2}
\end{equation}
and for the choice of operators
\begin{subequations}
\begin{eqnarray}
\hat{A}
&=&-\frac{X-X^{\dagger}}{2i}
= \sin \frac{2 \pi \hat{p}}{N} 
=   \hat{A}^{\dagger}  \; ,     
\label{A periodic}     \\
\hat{B}
&=&\frac{Z-Z^{\dagger}}{2i}
= \sin \frac{2 \pi \hat{q}}{N} 
= \hat{B}^{\dagger},
\label{B periodic}
\end{eqnarray}
\label{A,B periodic}
\end{subequations}
which are physical observables formed using the Schwinger operators $\hat{X}$ and $\hat{Z}$ defined in App. \ref{schwinger}.
The inequality (\ref{robertson 2}) requires single-valuedness of the observables
(see Ref. \cite{peres_book}, pp. 91-94, and Ref. \cite{ballentine}), a property which is fulfilled with the choice (\ref{A,B periodic}). 


For the quantities entering Eq. (\ref{robertson 2})
we find the following results (we made the choice $p^{(n_0)}=0$)
\begin{subequations}
\begin{eqnarray}
\langle\hat{A}\rangle_{| \psi^{(0, \delta p)}\rangle }
&=&   0
\label{<A>}         \\
\langle\hat{A}^2\rangle_{ |\psi^{(0, \delta p)}\rangle }
&=&\frac12 \left[
1-\frac{1}{\delta p +1} \frac{\sin (\frac{2\pi}{N}(\delta p +1))}{\sin \frac{2\pi}{N}}
\right]
\label{<A2>}  \\
\langle \hat{B}\rangle_{ |\psi^{(0, \delta p)} \rangle}
&=& 0
\label{B}  \\
\langle \hat{B}^2\rangle_{ |\psi^{(0, \delta p)} \rangle}
&=& \frac{1}{N(\delta p +1)}
\sum_{q=0}^{N-1}\sin^2 \frac{2\pi q}{N}
\left(
\frac{\sin\frac{ \pi q (\delta p +1)} {N} }{\sin \frac{\pi q}{N}}
\right)^2
\label{<B2>}  \\
\langle [\hat{A}, \hat{B}] \rangle_{ |\psi^{(0, \delta p)} \rangle}
&=& \frac{-i}{N(\delta p +1)}\sum_{q=0}^{N-1}
\sin \frac{2\pi q}{N} \;
 \frac{ \sin \frac{\pi q (\delta p +1) }{N} } {\sin \frac{\pi q}{N}}
\left[
\frac{\sin \frac{\pi(q-1)(\delta p+1)}{N}} {\sin \frac{\pi (q-1)}{N}} 
- \frac{\sin \frac{\pi(q+1)(\delta p+1)}{N}} {\sin \frac{\pi (q+1)}{N}} 
\right] 
\label{<[A,B]>}
\nonumber \\
\end{eqnarray}
\label{<A>,<B>,<A,B>}
\end{subequations}

For the following choices 
of $N$ and $\delta p$ satisfying Eqs. (\ref{dp, N}), 
Eqs. (\ref{<A>,<B>,<A,B>}) give the LHS and RHS of the inequality (\ref{robertson 2}) as
\begin{subequations}
\begin{eqnarray}
&& N=6, \;\;\delta p = 2, \hspace{1cm} L=0.3333, \;\;\;\; R=0.1667  \\
&& N=9, \;\;\delta p = 2, \hspace{1cm} L=0.3333, \;\;\;\; R=0.1667  \\
&& N=12, \;\;\delta p = 2, \hspace{1cm} L=0.3333, \;\;\;\; R=0.1667  \\
&& N=15, \;\;\delta p = 4, \hspace{1cm} L=0.2, \;\;\;\; R=0.0770  \\
&& N=20, \;\;\delta p = 4, \hspace{1cm} L=0.2, \;\;\;\; R=0.0784  \\
&& N=22, \;\;\delta p = 10, \hspace{1cm} L=0.0909, \;\;\;\; R=0.0162 ,
\end{eqnarray}
\end{subequations}
satisfying $L>R$.

\section{
Application to the successive measurement of momentum and position in the continuous case}
\label{application to continuous position-momentum}

In this section we apply the formalism 
that was developed above
to the successive measurement of momentum and position in the continuous case.
The two observables of Eqs. (\ref{sp repr A,B 0}) will be taken as
\begin{subequations}
\begin{eqnarray}
&&\hat{A}^{\delta a}
= \sum_n p^{(n)} \hat{\mathbb{P}}_{p^{(n)}}^{\delta p}
\equiv \hat{p}^{\delta p} ,
\;\;\; {\rm where} \;\;\;\
\hat{\mathbb{P}}_{p^{(n)}}^{\delta p}
= \int_{p^{(n)}-\delta p / 2}^{p^{(n)}+\delta p /2}dp' | p'\rangle \langle p' |,
\;\;\; p^{(n)} = n \delta p.
\label{p-Delta-p} \\
&&\hat{B}
= \int dx \; x \; \hat{\mathbb{P}}_{x}
\equiv \hat{x} ,  
\;\;\;\;\; {\rm where} \;\;\;\;\;
\hat{\mathbb{P}}_{x}
=  | x\rangle \langle x| \; .
\label{x}
\end{eqnarray}
\end{subequations}
Here, $\hat{p}^{\delta p}$ is a low-resolution version of the momentum operator, while,
just as before, the second observable, $\hat{x}$, is taken as the full-resolution position operator.
As in the previous sections, the various intervals 
$[p^{(n)}-\delta p / 2, p^{(n)} + \delta p / 2]$ in the definition of
$\hat{p}^{\delta p}$ are {\em disjoint}.
The projector $\hat{\mathbb{P}}_{p^{(n)}}^{\delta p}$ filters coherently the $p$ components inside an interval of size $\delta p$ centered at $p^{(n)}$;
the resulting $\hat{p}^{\delta p}$ is a discretized version of the momentum operator $\hat{p}$.
Properties (\ref{P-delta-a-property 1})-(\ref{P-delta-a-property 3})
translate to the present operators 
$\hat{\mathbb{P}}_{p^{(n)}}^{\delta p}$ as:

1) They are well defined projector operators, satisfying
\begin{equation}
\hat{\mathbb{P}}_{p^{(n)}}^{\delta p} \;
\hat{\mathbb{P}}_{p^{(n')}}^{\delta p}
= \delta_{n n'} \hat{\mathbb{P}}_{p^{(n)}}^{\delta p} \; ;
\end{equation}

2) They are eigen-projectors of the operator $\hat{p}^{\delta p}$; i.e.,
\begin{equation}
\hat{p}^{\delta p} \; \hat{\mathbb{P}}_{p^{(n)}}^{\delta p}
= p^{(n)} \hat{\mathbb{P}}_{p^{(n)}}^{\delta p} \; ;
\end{equation}

3) However, they are {\em not} eigen-projectors of the operator $\hat{p}$; i.e.,
\begin{equation}
\hat{p} \; \hat{\mathbb{P}}_{p^{(n)}}^{\delta p}
= \int_{p^{(n)}-\delta p / 2}^{p^{(n)}+\delta p /2} p' | p'\rangle \langle p' | dp'
\approx p^{(n)} \hat{\mathbb{P}}_{p^{(n)}}^{\delta p} \; ,
\end{equation}
the equality sign holding only approximately, if the interval $\delta p$
is small enough.

At $t=t_1$, the system momentum $\hat{p}^{\delta p}$ is measured, {\em followed} by a measurement of the system position at time $t_2$.
We shall assume that the initial state of the system is the pure state 
$\hat{\rho}=| \psi \rangle \langle \psi |$.
As usual, we shall only consider the case of strong coupling between the system and the first probe.

The conditional Wigner's formula of 
Eq. (\ref{cond_wigner_formula}) takes the form, 
from Eqs. (\ref{pert psi non-deg proj Da neq 0})-(\ref{N(bm),D}).
\begin{equation}
{\cal W}(x |p^{(0)} , \delta p)
= \frac{  
\langle \psi |
\hat{\mathbb{P}}_{p^{(0)}}^{\delta p}\; \hat{\mathbb{P}}_{x} \;
\hat{\mathbb{P}}_{p^{(0)}}^{\delta p} 
| \psi \rangle 
}
{\langle \psi |  \hat{\mathbb{P}}_{p^{(0)}}^{\delta p} | \psi \rangle
}  
=\frac{N(x)}{D} \; .
\label{cond_wigner_formula_posi_mom}
\end{equation}
This is the probability density to find the position value $x$, conditioned on having found the momentum in an interval $\delta p$ around the value $p^{(0)} $.
One finds the particular cases of Eqs. (\ref{N(bm),D})
\begin{subequations}
\begin{eqnarray}
N(x) &=&\frac{1}{2\pi}\left|\int_{p^{(0)}-\delta p/2}^{p^{(0)}+\delta p/2}dp \; \tilde{\psi}(p) \; {\rm e}^{ipx} \right|^2
\label{N} \\
D &=& \int_{p^{(0)}-\delta p/2}^{p^{(0)}+\delta p/2} dp  \; \left| \tilde{\psi}(p) \right|^2    
\label{D}
\end{eqnarray}
\label{N,D p,x}
\end{subequations}
Here, $\tilde{\psi}(p) $ is the wave function in momentum space.

\begin{figure}[h]
\centering
\includegraphics[width=10cm,height=10.0cm]
{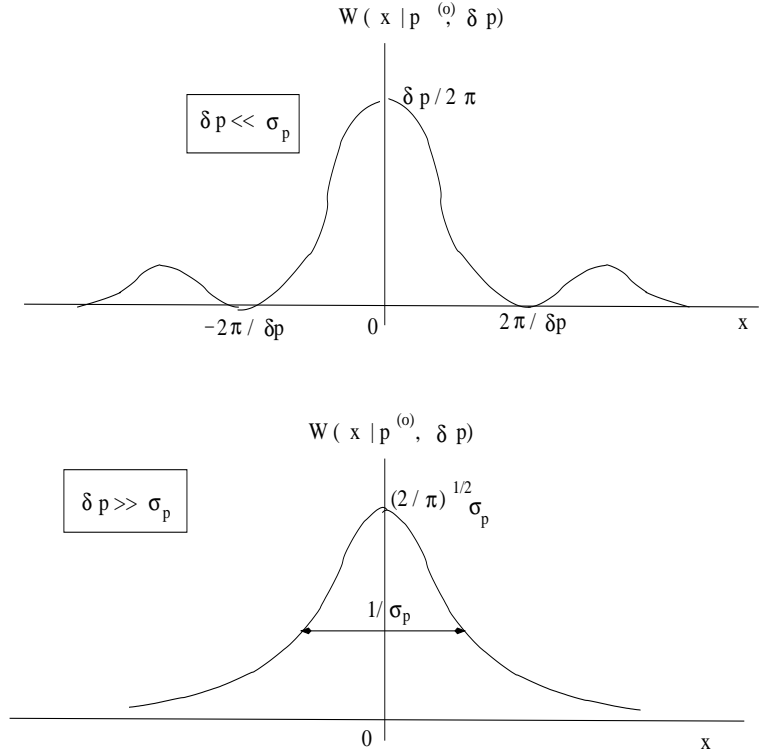}
\caption{
\footnotesize{The probability distribution, as given by 
Eqs. 
(\ref{cond_wigner_formula_posi_mom 2}), 
for the position $x$, conditioned on having found the momentum $p$ in an interval $\delta p$ around the value $p^{(0)}$, for the wave function of Eq. (\ref{psi(p)}) (schematic).
Upper curve: the case $\delta p \ll \sigma_p$;
lower curve: the case $\delta p \gg \sigma_p$.
}
}
\label{W(x|p)}
\end{figure}

As an illustration, consider the wave function
\begin{equation}
\tilde{\psi}(p)
=\frac{{\rm e}^{-\frac{(p-p^{(0)})^2}{4\sigma_p^2}}}
{(2\pi \sigma_p^2)^{1/4}},
\label{psi(p)}
\end{equation}
real and centered at $p^{(0)}$ for convenience,
as assumed in the paragraph following Eq. (\ref{N(bm),D}).
The quantities $N(x)$ and $D$ of Eqs. (\ref{N,D}) take the form
\begin{subequations}
\begin{eqnarray}
N(x) &=&\frac{1}{2\pi}\left|\int_{-\delta p/2}^{\delta p/2} dp' \;
\frac{{\rm e}^{-\frac{(p')^2}{4\sigma_p^2}}}
{(2\pi \sigma_p^2)^{1/4}} \;
{\rm e}^{ip'x} \right|^2
\approx \left\{
\begin{array}{cccc}
\frac{1}{2\pi \sqrt{2\pi \sigma_p^2}}
\left(
\frac{\sin \frac{x \delta p}{2}}{\frac{x}{2}}
\right)^2 \; ,  
& {\rm for} & 
\delta p \ll  \sigma_p, \; x \delta p = {\rm arbitr.}  \\
|\psi(x)|^2 \; , 
& {\rm for} &
\delta p \gg \sigma_p
\end{array}
\right.
\label{N gauss} \\
D &=& \int_{-\delta p/2}^{\delta p/2} dp'  \;  
\frac{{\rm e}^{-\frac{(p')^2}{2\sigma_p^2}}}
{\sqrt{2\pi \sigma_p^2}} 
\approx \left\{
\begin{array}{cccc}
\frac{\delta p}{\sqrt{2 \pi \sigma_p^2}} \; ,  
& {\rm for} & 
\delta p \ll  \sigma_p   \\
1 \; , 
& {\rm for} &
\delta p \gg \sigma_p
\end{array}
\right.
,
\label{D gauss}
\end{eqnarray}
\label{N,D gauss}
\end{subequations}
as particular cases of Eqs. (\ref{N(bm),D}) and (\ref{N,D}).
An analysis of the approximations involved in Eqs. (\ref{N,D gauss})
is presented in App. \ref{the approx N,D gauss}.
The conditional Wigner's formula, Eq. (\ref{cond_wigner_formula_posi_mom}),
then gives
(see Eqs. (\ref{cond_wigner_formula_an_bm}))
\begin{equation}
{\cal W}(x |p^{(0)} , \delta p)
=
\left\{
\begin{array}{ccccc}
\frac{1}{2 \pi \delta p}  \left[ \frac{\sin (\frac{x \delta p}{2})}{x/2}\right]^2 
&\Rightarrow  
& \frac{\delta x/2}{2} \sim \frac{\pi}{\delta p}, &
{\rm for} & \delta p \ll  \sigma_p   
\;\;\;\; (a) \\
|\psi(x)|^2  
&\Rightarrow
&\Delta x \equiv \sqrt{var x} = \frac{1}{2 \sigma_p}, &
{\rm for} &\delta p \gg  \sigma_p
\;\;\;\; (b)
\end{array}
\right.
\label{cond_wigner_formula_posi_mom 2}
\end{equation}
In the first case of Eq. (\ref{cond_wigner_formula_posi_mom 2}a), 
{\em the width in position of the $x$ distribution as a function of the resolution 
$\delta p$ constitutes the uncertainty relation} we are concerned with.
We measure the width according to Eq. (\ref{delta b}), since 
the second moment of $x$ diverges
due to the 
sharp cutoff in the integral over $p'$: 
$\delta x /2$ is defined as the position of the first zero of ${\cal W}(x |p^{(0)} , \delta p)$.
The result is that, having measured $\hat{p}$ with a resolution
$\delta p \ll \sigma_p$ around $p^{(0)}$
(see Fig. \ref{fig1}),
a successive measurement of $\hat{x}$ has a probability which is spread over an interval
$\delta x \sim \frac{4\pi}{\delta p}$, so that
\begin{equation}
\delta x  . \delta p \sim 4 \pi .
\label{cond_wigner_formula_posi_mom_cont}
\end{equation}
This is illustrated in Fig. \ref{W(x|p)} (a).

In the second case of Eq. (\ref{cond_wigner_formula_posi_mom 2}b),
having measured $\hat{p}$ with
a resolution
$\delta p \gg \sigma_p$ around 
$p^{(0)}$ (see Fig. \ref{fig2}),
a successive measurement of $\hat{x}$ has a probability which is spread over an interval
$\Delta x \sim \frac{1}{2 \sigma_p}$,
which is unrelated to the resolution $\delta p$;
here the usual variance is well defined.
This is illustrated in Fig. \ref{W(x|p)} (b).

The UR we have been contemplating looks for a 
{\em relation between the width of the $x$ distribution}, or conditional Wigner function, 
${\cal W}(x |p^{(0)} , \delta p)$, 
{\em and $\delta p$, the resolution in $p$ of the first measurement}, which is unrelated to the width $\sigma_p$ of the original wave function
$\tilde{\psi}(p)$.
That such an UR appears when $\delta p \ll \sigma_p$
is clear from the following considerations.

i) We may write the distribution 
${\cal W}(x |p^{(0)} , \delta p)$ of 
Eq. (\ref{cond_wigner_formula_posi_mom}) as
\begin{subequations}
\begin{eqnarray}
{\cal W}(x |p^{(0)} , \delta p)
&=& \frac{\left|
\int_{-\infty}^{\infty}
\widetilde{\psi}(p)\theta(p \in p^{(0)}, \delta p)
\frac{{\rm e}^{ipx}}{\sqrt{2\pi}} dp
\right|^2}
{
\int_{-\infty}^{\infty}
\left|\widetilde{\psi}(p)\theta(p \in p^{(0)}, \delta p)
\right|^2 dp
}  
\label{cond_wigner_formula_posi_mom 3 a} \\
&=& 
\left|
\int_{-\infty}^{\infty}
\widetilde{\psi}_{p^{(0)}, \delta p}^{(normalized)}(p)
\frac{{\rm e}^{ipx}}{\sqrt{2\pi}} dp
\right|^2 \; .
\label{cond_wigner_formula_posi_mom 3 b}  
\end{eqnarray}
If 
\begin{eqnarray}
\widetilde{\psi}_{p^{(0)}, \delta p}(p)
&=& \widetilde{\psi}(p)\theta(p \in p^{(0)}, \delta p)
\label{cond_wigner_formula_posi_mom 3 d}
\end{eqnarray}
\label{cond_wigner_formula_posi_mom 3} 
\end{subequations}
denotes the original system wave function $\widetilde{\psi}(p)$
``chopped off'' to the interval $(p^{(0)}-\delta p/2, p^{(0)}+\delta p/2)$
[with $\theta(p \in p^{(0)}, \delta p)=1$ if $p \in (p^{(0)}, \delta p)$ and $=0$ otherwise], 
then
\begin{eqnarray}
\widetilde{\psi}_{p^{(0)}, \delta p}^{(normalized)}(p)
&=& \frac{\widetilde{\psi}_{p^{(0)}, \delta p}(p)}
{(\widetilde{\psi}_{p^{(0)}, \delta p}(p),\widetilde{\psi}_{p^{(0)}, \delta p}(p))^{1/2}} \; .
\label{cond_wigner_formula_posi_mom 3 c} 
\end{eqnarray}

ii) We recall again that 
$\delta p$ is the resolution in $p$ of the first measurement, which is unrelated to the width $\sigma_p$ of the original wave function
$\tilde{\psi}(p)$.
If $\delta p \ll \sigma_p$, as in 
Eq. (\ref{cond_wigner_formula_posi_mom 2}a),
the effective width of the resulting wave function 
$\widetilde{\psi}_{p^{(0)}, \delta p}^{(normalized)}(p)$
is $\delta p$;
if $\delta p \gg \sigma_p$, as in 
Eq. (\ref{cond_wigner_formula_posi_mom 2}b),
the effective width of the resulting wave function is $\sigma_p$.

iii) Since we have in (\ref{cond_wigner_formula_posi_mom 3 b}) the Fourier transform of $\widetilde{\psi}_{p^{(0)}, \delta p}^{(normalized)}(p)$, we expect, when 
$\delta p \ll \sigma_p$, an UR as in 
Eq. (\ref{cond_wigner_formula_posi_mom 2}a), where $\delta x$ is inversely proportional to the resolution $\delta p$, while when $\delta p \gg \sigma_p$, we have 
Eq. (\ref{cond_wigner_formula_posi_mom 2}b), where $\Delta x$ is inversely proportional to $\sigma_p$, unrelated to the resolution $\delta p$.

\section{Summary and Conclusions}
\label{conclu}

In the present paper we investigated the quantum-mechanical uncertainty relation (UR) arising from the successive measurement of two observables carried out on the {\em same system}.
This formulation is 
closer to the original Heisenberg conception of the UR,
and has to be contrasted with the familiar textbook formulation, in which,
when the two observables are position and momentum, one contemplates the standard deviation of the position over an ensemble of systems at time $t$, vs the standard
deviation of the momentum over an independent ensemble of equally prepared systems, also at time 
$t$.

We employed an extension of the von Neumann model of measurement, in which
two probes interact with the same system at two successive times $t_1$ and $t_2$, 
so we can exhibit explicitly how the disturbing effect of the first interaction affects the second measurement.
The first interaction is designed to measure the observable $\hat{A}^{\delta a}$, i.e., $\hat{A}$ with {\em resolution} $\delta a$, and the second interaction measures the observable $\hat{B}$.
For simplicity, we have assumed in all cases that the 
spectra of the two observables are {\em non-degenerate}.

At time $t_f > t_2$, i.e., after the system-probes interactions are over, we detect, for an {\em individual system} $s$,
the dynamical variables $Q_1, Q_2$
pertaining to the {\em two probes}: 
this detection can be realized, since the dynamical variables 
$Q_1, Q_2$ commute.
From such a detection, we uncover information on the 
{\em system proper}.


We found that in the limit of {\em strong coupling} between the system and the first probe, detecting the statistical distribution of the probe variable $Q_2$ conditioned on $Q_1$, specifically
$p(Q_2|Q_1=\epsilon_1 a ^{(n_0)})$,
we obtain information on the {\em statistical distribution}
${\cal W}(b_m|a^{(n_0)}, \delta a)$
of the {\em system variable} $b_m$ 
{\em conditioned} on having found $a_n$
around the value $a^{(n_0)}$ in a measurement 
with {\em resolution} $\delta a$. 
The distribution ${\cal W}(b_m|a^{(n_0)}, \delta a)$ is given by Wigner's formula (\ref{cond_wigner_formula});
its width 
as function of the resolution
$\delta a$ of the first measurement constitutes the 
{\em uncertainty relation for successive measurements} of main interest in the present paper.
When the second moment of this distribution exists, we could express in a very general way the UR for successive measurements as the inequality (\ref{robertson}),
involving the commutator of the two observables measured successively.

We illustrated the UR in the case of
the successive measurement of the {\em position}- and {\em momentum-like} operators defined for Schwinger's model in a discrete, finite-dimensional Hilbert space, giving the uncertainty relation of 
Eq. (\ref{Deltap.Deltaq ge N}).
We verified the validity of the inequality 
(\ref{robertson}) for this case. 

We also illustrated this relation for the case of the successive measurement of 
{\em position and momentum in the continuous case}.
This case allows understanding various features of the UR 
that was developed in the present paper.
We found the UR
(\ref{cond_wigner_formula_posi_mom_cont}), that we reproduce here,
\begin{equation}
\delta x  . \delta p \sim 4 \pi ,
\label{cond_wigner_formula_posi_mom_cont 1}
\end{equation}
between the {\em width $\delta x$ of the $x$ distribution}
or conditional Wigner function, 
${\cal W}(x |p^{(0)} , \delta p)$
of Eq. (\ref{cond_wigner_formula_posi_mom}), 
and the {\em resolution $\delta p$ of the first measurement}.
This relation was found in the situation when $\delta p \ll \sigma_p$, $\sigma_p$ being the width of the original system wave function 
$\tilde{\psi}(p)$, which is unrelated to $\delta p$.
We could give a clear explanation why for this UR to occur we need $\delta p \ll \sigma_p$.
The sharp cutoff appearing in Eqs. (\ref{cond_wigner_formula_posi_mom 3})
is responsible for the divergence of the second moment of 
${\cal W}(x |p^{(0)} , \delta p)$.

When the two observables commute, 
$[\hat{A}, \hat{B}]=0$, we found $\delta b = \delta a$, 
as described in Eqs. 
(\ref{db=da for A=B}) and  (\ref{db=da for a neq B [A,B]=0}).
For a more general value of the commutator $[\hat{A}, \hat{B}]$ we found, in a model described in Sec. \ref{[A,B] arbitr}, the result of Eqs. (\ref{DxDx' 2})
and (\ref{inequalities}).
The equality, Eq. (\ref{DxDx' 2}), contains the commutator of the two measured observables and depends on $\delta x^2$, $(\delta x')^2$ and $(\delta x')^4$.
The inequality (\ref{inequality a}) is similar to the standard Robertson inequality
[\onlinecite{robertson}], except that here there is no dependence left on the original system wave function, due to the assumption 
$\delta x' \ll \sigma_{x'}$.


As already noted, in the present paper we exhibited the UR associated with the successive measurement of two observables when the spectra of these observables are non degenerate. 
Extension of the analysis to observables with degenerate spectra is left for future studies.

\acknowledgments

P. A. M. acknowledges support from the Sistema Nacional de Investigadores, Mexico, and from Conacyt, Mexico, under contract No. 282927.
He is also grateful to M. Bauer, for making him aware of some of the articles in the early literature on the subject, as well as of some recent ones.

\appendix

\section{Derivation of Eq. (\ref{p(Q2|Q1)})}
\label{derivation p(Q2|Q1)} 

The unitary evolution operator associated with the Hamiltonian
of Eq. (\ref{V 2 probes}) satisfies the equations
\begin{subequations}
\begin{eqnarray}
i\frac{\partial \hat{U}}{\partial t} 
&=& \hat{H} \hat{U}
\\
\hat{U}(0)&=& I \; ,
\label{idU/dt}
\end{eqnarray}
\end{subequations}
in units of $\hbar=1$.
By direct substitution one verifies that the result is given by \cite{johansen_mello_2008,mello_lasp_aip_2014}
\begin{subequations}
\begin{eqnarray}
 \hat{U}(t) =e^{-i\int_0^t \epsilon_2 g_2(t')\hat{B} \hat{P}_2 \; dt'}
 e^{-i\int_0^t\epsilon_1 g_1(t')\hat{A}^{\delta a} \hat{P}_1 \; dt'}
\label{U 1}
 \\
 =e^{-i\epsilon_2 G_2(t) \hat{B} \hat{P}_2}
 e^{-i\epsilon_1 G_1(t) \hat{A}^{\delta a} \hat{P}_1},
 \label{U 2}
\end{eqnarray}
\label{U}
\end{subequations}
where we have defined
\begin{subequations}
\begin{eqnarray}
\int_{0}^{t} g(t')dt' &\equiv& {G(t)}   \\
G(0) &=& 0, \;\;\;  G(\infty) =1
\end{eqnarray}
\end{subequations}

The state of the system plus the two probes, $\pi_1$ , $\pi_2$, is described, at $t=0$, by the density matrix
\begin{equation}
\hat{\rho}_0 = \hat{\rho}_{s} \; \hat{\rho}_{\pi_1} \; \hat{\rho}_{\pi_2}.
\label{rho t0 12}
\end{equation}
At later times, the density matrix is given by
\begin{subequations}
\begin{eqnarray}
\hat{\rho}(t)
&=& \hat{U}(t) \hat{\rho}_0 \hat{U}^{\dagger}(t)  \\
&=&e^{-i\epsilon_2 G_2(t) \hat{B} \hat{P}_2}
 e^{-i\epsilon_1 G_1(t) \hat{A}^{\delta a} \hat{P}_1}
\hat{\rho}_{s} \; \hat{\rho}_{\pi_1} \; \hat{\rho}_{\pi_2} 
e^{i\epsilon_1 G_1(t) \hat{A}^{\delta a} \hat{P}_1}
 e^{i\epsilon_2 G_2(t) \hat{B} \hat{P}_2}
  \\
&=&
\sum_{nn'mm'}
e^{-i\epsilon_2 G_2(t) \hat{B} \hat{P}_2}
\hat{\mathbb{P}}_{b_m} e^{-i\epsilon_1 G_1(t) \hat{A}^{\delta a} \hat{P}_1}
\hat{\mathbb{P}}_{a^{(n)}}^{\delta a} \hat{\rho}_{s} \; \hat{\rho}_{\pi_1} \; \hat{\rho}_{\pi_2} 
\hat{\mathbb{P}}_{a^{(n')}}^{\delta a} e^{i\epsilon_1 G_1(t) \hat{A}^{\delta a} \hat{P}_1}
\hat{\mathbb{P}}_{b_{m'}} e^{i\epsilon_2 G_2(t) \hat{B} \hat{P}_2} 
\nonumber \\
&=&
\sum_{nn'mm'}
\left(\hat{\mathbb{P}}_{b_{m}}  \hat{\mathbb{P}}^{\delta a}_{a^{(n)}}  \hat{\rho}_{s}  \hat{\mathbb{P}}^{\delta a}_{a^{(n')}} \hat{\mathbb{P}}_{b_{m'}}\right)
\nonumber \\
&&\hspace{1cm}\times \left[
e^{-i\epsilon_1 G_1(t) a^{(n)} \hat{P_1}} \hat{\rho}_{\pi_1}
e^{i\epsilon_1  G_1(t) a^{(n')}\hat{P_1}} \right]
\left[
e^{-i\epsilon_2  G_2(t) b_m \hat{P_2}} \hat{\rho}_{\pi_2}
e^{i\epsilon_2   G_2(t) b_{m'}\hat{P_2}}\right] 
\end{eqnarray}
\label{rho t>0}
\end{subequations}

For $t \gg t_2$, i.e., after 
the second interaction has ceased to act, we have ($f$ stands for ``final")
\begin{eqnarray}
\hat{\rho}_f \equiv \hat{\rho}_{t_2<t}
&=& \sum_{nn'mm'}
\left(
\hat{\mathbb{P}}_{b_{m}} \hat{\mathbb{P}}_{a^{(n)}}^{\delta a}  \hat{\rho}_{s}  \hat{\mathbb{P}}_{a^{(n')}}^{\delta a} \hat{\mathbb{P}}_{b_{m'}}
\right)
\nonumber \\
&&\hspace{1cm}\times 
\left(
e^{-i\epsilon_1 a^{(n)} \hat{P_1}} \hat{\rho}_{\pi_1}
e^{ i\epsilon_1 a^{(n')}\hat{P_1} } 
\right)
\left(
e^{-i\epsilon_2 b_m \hat{P_2}} \hat{\rho}_{\pi_2}
e^{i\epsilon_2 b_{m'}\hat{P_2}}
\right) .
\label{rho f}
\end{eqnarray}

From Eq. (\ref{rho f}) we find the final joint probability density of the two 
{\em commuting} position observables $\hat{Q}_1$, $\hat{Q}_2$
in terms of the position projectors $\mathbb{P}_{Q_1}$
and $\mathbb{P}_{Q_2}$ as
\begin{eqnarray}
p_{f}(Q_1, Q_2) 
&=& {\rm Tr}(\hat{\rho}_{f} \hat{\mathbb{P}}_{Q_1} \hat{\mathbb{P}}_{Q_2})
\\
&=& \sum_{n,n',m}{\rm Tr}_s 
(\hat{\rho}_{s}\mathbb{\hat{P}}^{\delta a}_{a^{(n')}}
\mathbb{\hat{P}}_{b_{m}}
\mathbb{\hat{P}}^{\delta a}_{a^{(n)}})
\nonumber \\
&& \hspace {1cm}
\times \chi_1(Q_1-\epsilon_1 a^{(n)}) 
\chi_1^{*}(Q_1-\epsilon_1 a^{(n')})
\big|\chi_2(Q_2 - \epsilon_2 b_m)\big|^2
\label{p(Q1,Q2) 1}
\end{eqnarray}
where we have assumed pure states $\chi_i(Q_i)$, $i=1,2$, for the two probes at $t=0$.
To be specific, we assume, for the probe wave functions, the Gaussian model
\begin{equation}\chi_{i}(Q_i)
=\frac{ e^{-\frac{Q_i^2}{4\sigma^2_{Q_i}}} }
{(2\pi\sigma^2_{Q_i})^{1/4}} .
 \label{gaussian-model}
\end{equation}
The joint probability density (\ref{p(Q1,Q2) 1}) then becomes
\begin{eqnarray}
p_f(Q_1,Q_2)
 = \sum_m \Bigg\{
 \sum_{n,n'}{\rm Tr}_s (\hat{\rho}_{s}
\hat{\mathbb{P}}^{\delta a}_{a^{(n')}}\hat{\mathbb{P}}_{b_{m}}
\hat{\mathbb{P}}^{\delta a}_{a_{(n)}} )              )
g_{nn'}(\epsilon_1/\sigma_{Q_1})
\frac
{
 e^{-\frac{\big(Q_1-\epsilon_1\frac{a^{(n)} + a^{(n')}}{2}\big)^2}{2\sigma_{Q_1}^2}}}
 {\sqrt{2\pi \sigma_{Q_1}^2}}
 \Bigg\}\;
\frac {e^{-\frac{(Q_2-\epsilon_2 b_m )^2}{2\sigma_{Q_2}^2}} } {\sqrt{2\pi
\sigma_{Q_2}^2}} ,
\nonumber \\
 \label{p(Q1,Q2) 2}
\end{eqnarray}
where $g_{nn'}(\epsilon_1/\sigma_{Q_1})$ is given in Eq. (\ref{g}).

The final, marginal probability density of $Q_1$ is obtained integrating 
$p_f(Q_1,Q_2)$ over $Q_2$, with the result
\begin{equation}
p_f(Q_1)
= \sum_n {\rm Tr}_s (\hat{\rho}_s \hat{\mathbb{P}}^{\delta a}_{a^{(n)}}) \;
\frac{{\rm e}^{-\frac{\left(Q_1 - \epsilon_1 a^{(n)}\right)^2}{2 \sigma_{Q_1}^2}}}{\sqrt{2 \pi \sigma_{Q_1}^2}}.
\label{pf(Q1)}
\end{equation}
Notice that this last equation can be obtained from Eq. (7)
of Ref. [\onlinecite{johansen_mello_2008}],
which describes a 
{\em single measurement for the full resolution} case $\delta a = 0$, with the replacements
\begin{equation}
a_n \Rightarrow a^{(n)}, \;\;\;\;\; 
\hat{\mathbb{P}}_{a_n} \Rightarrow
\hat{\mathbb{P}}^{\delta a}_{a^{(n)}} \; .
\end{equation}
Eq. (\ref{pf(Q1)}) describes a 
{\em single measurement for the  low-resolution} case 
$\delta a \neq 0$.

The $Q_2$ probability density, conditioned on a given value of $Q_1$, is the ratio
\begin{equation}
p_f(Q_2|Q_1)
= \frac{p_f(Q_1,Q_2)}{p_f(Q_1)}
\label{p(Q2|Q1) 0}
\end{equation}
Substituting the results of Eqs. (\ref{p(Q1,Q2) 2}) and (\ref{pf(Q1)}),
we obtain Eq. (\ref{p(Q2|Q1)}) of the text.

\section{The Schwinger operators}
\label{schwinger}

We consider an $N$-dimensional Hilbert space spanned by $N$ distinct states 
$|q\rangle$, with $q=0,1, \cdots ,(N-1)$, which are subject to the periodic condition
$|q+N\rangle=|q\rangle$.
These states are designated as the ``reference basis" of the space.  
We follow Schwinger \cite{schwinger} and introduce the unitary operators $\hat{X}$ and $\hat{Z}$, defined by their action on the states of the reference basis by the equations
\begin{subequations}
\begin{eqnarray}
\hat{Z}|q\rangle
&=&\omega^q|q\rangle, \;\;\;\; \omega=e^{2 \pi i/N},
\label{Z}  \\
\hat{X}|q\rangle &=& |q+1\rangle .
\label{X}
\end{eqnarray}
\label{Z,X}
\end{subequations}
The operators $\hat{X}$ and $\hat{Z}$ fulfill the periodic condition
\begin{equation}
\hat{X}^N = \hat{Z}^N = 
\hat{\mathbb{I}},
\label{X,Z periodic}
\end{equation}
$\hat{\mathbb{I}}$ being the unit operator.
These definitions lead to the commutation relation
\begin{equation}
\hat{Z}\hat{X}=\omega \hat{X}\hat{Z} .
\label{comm Z,X}
\end{equation}
The two operators $\hat{Z}$ and $\hat{X}$ form a complete algebraic set, in that only a multiple of the identity commutes with both \cite{schwinger}. 
As a consequence, any operator defined in our $N$-dimensional Hilbert space can be written as a function of $\hat{Z}$ and $\hat{X}$.

We introduce the Hermitean operators  $\hat{p}$ and $\hat{q}$,
which play the role of ``momentum-like" and ``position-like" operators, through the equations
\cite{de_la_torre-goyeneche,durt_et_al}
\begin{subequations}
\begin{eqnarray}
\hat{X}
&=& \omega^{-\hat{p}}
= e^{-\frac{2\pi i}{N}\hat{p}} \; ,
\label{X(p)} \\
\hat{Z}
&=& \omega^{\hat{q}}
=e^{\frac{2\pi i}{N}\hat{q}} \; .
\label{Z(q)} 
\end{eqnarray}
\label{X(p),Z(q)}
\end{subequations}
What we defined as the reference basis can thus be considered as the ``position basis".
With (\ref{comm Z,X}) and definitions (\ref{X(p),Z(q)}), the commutator of $\hat{q}$ and $\hat{p}$ in the continuous limit \cite{de_la_torre-goyeneche,durt_et_al}
is the standard one, $[\hat{q},\hat{p}]=i$.

\section{The approximations involved in Eqs. (\ref{N,D gauss})}
\label{the approx N,D gauss}

1) We first consider the quantity $D$ of Eq. (\ref{D gauss}) in the limit 
$\delta p \ll \sigma_p$.
Setting $\sigma_p=1$ for simplicity, we can write
\begin{subequations}
\begin{eqnarray}
D &=& \frac{1}{\sqrt{2\pi}}D'   
\label{D' a}     \\
D' &=& \int_{-\delta p /2}^{\delta p /2} 
{\rm e}^{-\frac{p^2}{2}} dp 
\label{D' b}         \\
&=& \sqrt{2 \pi} \; {\rm erf}\left(\frac{\delta p}{2\sqrt{2}}\right)
\label{D' c}
\end{eqnarray}
\label{D'}
\end{subequations}
We expand the above in powers of $\delta p$ in two ways, in order to provide a check:

\noindent
a) We expand the error function in Eq. (\ref{D' c}) in powers of its argument, to obtain
\begin{eqnarray}
D' = \delta p -\frac13 \left(\frac{\delta p}{2}\right)^3 + \cdots
\end{eqnarray}
\noindent
b) We expand the exponential in  Eq. (\ref{D' b}) in powers of $p$ to obtain
\begin{eqnarray}
D' &=& \int_{-\delta p /2}^{\delta p /2} 
\left(1 - \frac{p^2}{2}+ \cdots \right) dp  \\
&=& \delta p -\frac13 \left(\frac{\delta p}{2}\right)^3 + \cdots
\end{eqnarray}
The first term is the result appearing in the first Eq. (\ref{D gauss}).

2)  We now turn to $N$ of Eq. (\ref{N gauss}), again in the limit 
$\delta p \ll 1$.
We write
\begin{subequations}
\begin{eqnarray}
N &=& \frac{1}{2\pi}\left[\frac{N'}{(2\pi)^{1/4}}\right]^2
\label{N' a}     \\
N'(x, \delta p) &=& \int_{-\delta p /2}^{\delta p /2} 
{\rm e}^{-\frac{p^2}{4}} {\rm e}^{ipx} dp 
\label{N' b}        
\end{eqnarray}
\label{N'}
\end{subequations}
We expand the Gaussian in  Eq. (\ref{N' b}) in powers of $p$ to obtain
\begin{subequations}
\begin{eqnarray}
N'(x, \delta p) 
&=& \int_{-\delta p /2}^{\delta p /2} 
\left(1 - \frac{p^2}{4}+ \cdots \right) {\rm e}^{ipx} dp 
\equiv  N'_0(x, \delta p)  + N'_2(x, \delta p) + \cdots
\label{N'(x,dp) a}   \\
&=& \frac{ \sin \frac{x \delta p}{2} }{\frac{x}{2}}
\left[
1-\frac14 \left(\frac{\delta p}{2}\right)^2 + \cdots 
\right]
+ \frac{1}{2x^2} 
\left[ 
\frac{\sin \frac{x \delta p}{2}}{\frac{x}{2}} 
- (\delta p) \cos \frac{x \delta p}{2} 
\right] + \cdots 
\nonumber \\
&& \hspace{10cm}
\delta p \ll 1, \;\;\;\;\;\; x \delta p = {\rm arbitrary}
\nonumber \\
\label{N'(x,dp) b}   
\end{eqnarray}
We call
\begin{eqnarray}
 N'_0(x, \delta p)
&=& \frac{ \sin \frac{x \delta p}{2} }{\frac{x}{2}}
\label{N'(x,dp) c}
\end{eqnarray}
\end{subequations}
the result used in the text,  Eq. (\ref{N gauss}).

Defining $z=\frac{x \delta p}{2}$, we can write
\begin{subequations}
\begin{eqnarray}
N'(z, \delta p) 
&=& \delta p 
\left\{
\frac{\sin z}{z} \left[1- \frac{(\delta p)^2}{16} + \cdots \right]
+ \frac{1}{8 z^2}
\left(\frac{\sin z}{z} -\cos z \right) (\delta p)^2 + \cdots 
\right\} \; ,
\nonumber \\
&& \hspace{10cm}
\delta p \ll 1, \;\;\;\;\; z \; {\rm arbitrary} \; ,
\nonumber \\ 
\label{N'(z,dp) a} \\
 N'_0(z, \delta p)
&=& \delta p \; \frac{\sin z}{z} .
\label{N'(z,dp) b} 
\end{eqnarray}
\end{subequations}
We now compare (\ref{N'(z,dp) a}) and (\ref{N'(z,dp) b}) for various values of $z$:
\begin{subequations}
\begin{eqnarray}
N'(z=0, \delta p) 
&=& \delta p \left[1- \frac{(\delta p)^2}{48} + \cdots \right],
\hspace{30mm}
N_0'(z=0, \delta p) = \delta p
\\
N'(z\ll 1, \delta p) 
&=& \delta p \left[1- \frac{(\delta p)^2}{48} - \frac{z^2}{6} 
+ \cdots \right], 
\hspace{20mm}
N'_0(z \ll 1, \delta p) 
= \delta p \left(1-\frac{z^2}{6}+ \cdots \right)
\nonumber \\ \\
N'(z= \pi /2, \delta p) 
&=& \frac{2}{\pi} \delta p 
\left[1-\left(\frac{1}{16}-\frac{1}{2\pi^2}\right)
(\delta p)^2  \cdots \right] ,
\hspace{8mm}
N'_0(z= \pi /2, \delta p)
= \delta p \; \frac{2}{\pi}
\\
N'(z= \pi, \delta p) 
&=& (\delta p)\frac{(\delta p)^2}{8 \pi^2}\; ,
\hspace{50mm}
N'_0(z= \pi, \delta p) = 0
\end{eqnarray}
\end{subequations}
We conclude that $N'_0(z, \delta p)$ differs from the next approximation 
$N'(z, \delta p)$ by a correction $O(\delta p)^3$.



\end{document}